\documentclass[12pt]{article}
\usepackage{amssymb,amsmath,amsthm,mathtools}

\usepackage{authblk,array,geometry} 
\usepackage{setspace}

\usepackage{bm,paralist,dsfont,nicefrac,bbm,booktabs}

\usepackage{etoolbox,algorithm}
\usepackage[noend]{algpseudocode}

\usepackage{tikz,float,tabularx,thm-restate}
\usetikzlibrary{shapes,arrows,patterns,decorations.pathreplacing}

\usepackage[driverfallback=hypertex,pagebackref=true,colorlinks]{hyperref}
\hypersetup{linkcolor=[rgb]{.7,0,0}}
\hypersetup{citecolor=[rgb]{0,.7,0}}
\hypersetup{urlcolor=[rgb]{.7,0,.7}}

\usepackage{xcolor} 

\usepackage[capitalize, nameinlink, noabbrev]{cleveref} 

\makeatletter
\newcommand{\algmargin}{\the\ALG@thistlm}
\algnewcommand{\parState}[1]{\State%
  \parbox[t]{\dimexpr\linewidth-\algmargin}{\strut #1\strut}}

\newtheorem{theorem}{Theorem}[section]
\newtheorem*{theorem*}{Theorem}

\newtheorem*{definition*}{Definition}

\newtheorem{lemma}[theorem]{Lemma}
\newtheorem*{lemma*}{Lemma}

\newtheorem{claim}[theorem]{Claim}
\newtheorem*{claim*}{Claim}

\newtheorem*{fact*}{Fact}

\newtheorem*{observation*}{Observation}

\newtheorem*{conjecture*}{Conjecture}

\newtheorem{corollary}[theorem]{Corollary}
\newtheorem*{corollary*}{Corollary}

\newtheorem*{remark*}{Remark}

\newtheorem{proposition}[theorem]{Proposition}
\newtheorem*{proposition*}{Proposition}

\newtheorem{example}[theorem]{Example}
\newtheorem*{example*}{Example}

\crefname{claim}{Claim}{Claims}

\usepackage[draft,inline,marginclue,index]{fixme}  
\fxsetup{inline,theme=color,mode=multiuser}
\FXRegisterAuthor{rrs}{arrs}{\colorbox{orange!90!yellow}{\hyperref[fixmelist]{\color{white}RRS}}}  
\FXRegisterAuthor{sp}{asp}{\colorbox{red}{\hyperref[fixmelist]{\color{white}SP}}} 
\FXRegisterAuthor{ub}{aub}{\colorbox{purple}{\hyperref[fixmelist]{\color{white}UB}}} 

\allowdisplaybreaks

\DeclarePairedDelimiter{\paren}{\lparen}{\rparen}

\DeclarePairedDelimiter{\bracket}{[}{]}
\DeclarePairedDelimiter{\set}{\lbrace}{\rbrace}

\DeclarePairedDelimiter{\abs}{\lvert}{\rvert}

\DeclareMathOperator*{\argmax}{arg\,max}

\DeclareMathOperator*{\poly}{poly}
\DeclareMathOperator*{\stab}{stab}

\def\E{\mathop{{}\mathbb{E}}}

\def\1{\mathds{1}}

\def\sw{\mathsf{SW}}
\def\opt{\mathsf{OPT}}

\def\mcI{\mathcal{I}}
\def\util{\mathsf{util}}
\def\utilag{\util^{ \mathsf{Ag} }}
\def\utilpr{\util^{ \mathsf{Pr} }}

\title{Contracts with Cooperating Principals}

\author{Umang Bhaskar\thanks{\href{mailto:umang@tifr.res.in}{umang@tifr.res.in}}}
\author{Soumyajit Pyne\thanks{\href{mailto:soumyajit.pyne@tifr.res.in}{soumyajit.pyne@tifr.res.in}}}
\author{Raghuvansh R. Saxena\thanks{\href{mailto:raghuvansh.saxena@tifr.res.in}{raghuvansh.saxena@tifr.res.in}}}

\affil{Tata Institute of Fundamental Research, Mumbai}

\date{}

\begin{document}

\maketitle             

\maketitle
\thispagestyle{empty}
\addtocounter{page}{-1}

\begin{abstract}
We introduce a novel contract design setting involving multiple principals, with the goal of studying cooperation between principals. In our setting, each principal has a separate contract design instance with the same agent, and the agent can take different actions across instances. The principals then share the resulting outcomes with each other and use all outcomes to design a joint contract that maximizes the total utility of all the principals.

We show that observing all outcomes gives the principals a much larger space of possible contracts, which can benefit them immensely. Specifically, we show that even two identical cooperating principals can have utility roughly $n$ times larger than is possible without cooperation, but no larger, where $n$ is the size of the agent's action set. Interestingly, this is not a worst-case phenomenon: As the number $k$ of identical principals grows, they can extract almost the entire social welfare in all contract design instances. Moreover, if each individual instance has bounded size, the optimal joint contract can be computed in time polynomial in $k$. However, this large space of contracts comes with complexity-theoretic challenges. We show that, for non-identical instances, as the number of principals grows, finding the optimal contract becomes $\mathrm{NP}$-hard, even when each individual instance has only three actions and three outcomes.

 In order to reduce the space of possible contracts, we control the degree of cooperation between principals by defining nearly-separable contracts. These are defined in terms of a cooperation parameter that restricts how much a principal's payment can differ based on the other principals' outcomes. When this difference is quantified additively, we show that cooperation provides very limited benefits. Perhaps surprisingly, quantifying the difference multiplicatively allows them to achieve improvements similar to the unrestricted case.

\end{abstract}

\newpage
\pagenumbering{roman} 
\tableofcontents

\newpage

\listoffixmes 

\newpage

\setcounter{page}{1}
\pagenumbering{arabic}

\section{Introduction}
\label{sec:intro}

Contract theory studies the interaction between a principal and an agent in which the principal (she) wants to incentivize the agent (he) to take an action beneficial to her, but is unable to observe or control the action taken. All the principal can observe is an outcome drawn from a distribution associated with the action, and each outcome gives her a corresponding reward. The goal of the principal is to design a contract specifying an outcome-based payment to the agent that maximizes her utility, which is her expected reward minus the expected payment, assuming the agent chooses an action maximizing his expected payment less the cost of the action, if any.

This hidden-action principal-agent model \cite{Ros73,Holmstrom79,GrossmanH83,Mir99} has widespread applications and a rapidly growing literature, with a recent survey in \cite{DFT-survey}. In this work, we further develop this theory by introducing contract design with multiple cooperating principals. Specifically, we consider the interaction between an agent and multiple principals, each with her own set of actions and outcomes, where the principals can observe and offer payments based on each other's outcomes with the goal of maximizing their total utility.

Examples of such settings abound: Imagine different departments within an organization contracting with a single web services provider for their webpages. By sharing the observable metrics with each other, these departments can incentivize the web service provider to provide a better service to the organization as a whole. One can also consider settings where the principals are motivated to cooperate due to a common interest. For example, hospitals in a city may jointly contract an ambulance provider for transporting patients, or schools in a city could jointly contract with a private bus service for providing transportation to students, with a common goal of maximizing safety and timeliness.

\subsection{Our Contributions}
\label{sec:intro:contribution}

We view the model for cooperating principals as one of our significant contributions. Prior work has explored instances with multiple principals through the common agency setting \cite{bernheim-1} which is better suited to principals with competing interests. In contrast, our model is designed to study cooperation between principals, who share their outcomes with each other with the goal of maximizing their total utility. 

\paragraph{The complexity of cooperation.} With cooperating principals that share their outcomes, the first question one might ask is how this differs from the case of just a single principal with access to all the outcomes. As the number of possible outcomes with $k$ principals increases exponentially with $k$, this amounts to asking whether modeling the principals as $k$ separate instances gives us any advantage over modeling them as a single, exponentially larger instance. For the case of identical instances that are not too large individually, we show that the answer is yes.

\begin{theorem}[Informal; formal version in \cref{thm:identical-polytime}]
\label{thm:identical-polytime-informal}
An optimal contract for $k$ identical principals, each with $n$ actions and $m$ outcomes, interacting with a common agent can be computed in time $k^{ O\paren*{ nm + n^2 } } \cdot \poly(L)$ where $L$ is the bit-complexity of the single-principal instance.
\end{theorem}
 
In contrast, modeling the $k$ principals as a single instance would result in an instance where the number of actions and the number of outcomes scale exponentially with $k$. This would give the same exponential in $k$ dependence on the running time to find the optimal contract, which, as $k$ grows, is much worse than \cref{thm:identical-polytime-informal}. Indeed, the message of \cref{thm:identical-polytime-informal} is that treating many copies of the same instance as a single, much larger instance loses the inherent structure that is fundamental to the problem. Our model captures this structure explicitly and is therefore better suited.

To prove \cref{thm:identical-polytime-informal}, we model the task of finding the optimal contract as solving a set of linear programs. Crucially, the fact that the instances are identical allows us to identify certain symmetries in the action and outcome profiles of the agent with respect to the $k$ principals. Due to the symmetries, we can get better bounds on the complexity of the linear programs, ultimately giving \cref{thm:identical-polytime-informal}. 
We note that \cref{thm:identical-polytime-informal} outputs the optimal contract succinctly in terms of these symmetries, instead of outputting a payment for every possible tuple of $k$ outcomes. This is necessary as otherwise the time required to describe the optimal contract itself would be too large.

We believe similar bounds can be obtained even for non-identical instances, provided the total number of distinct instances is small. In sharp contrast, for general, non-identical instances, we show that computing the optimal contract is $\mathrm{NP}$-hard.

\begin{theorem}[Informal; formal version in \cref{thm:non-identical-np-hard}]
\label{thm:non-identical-np-hard-informal}
An optimal contract for $k$ non-identical principals is $\mathrm{NP}$-hard to compute, even if each principal has three outcomes and three actions, and the optimal contract is supported on exactly one (joint) outcome.
\end{theorem}

Thus, the hardness in \cref{thm:non-identical-np-hard-informal} does not stem from the contract being hard to describe, or from the fact that each individual instance is hard to solve. It arises from the intricate nature of the cooperation required to achieve the optimal utility. We show \cref{thm:non-identical-np-hard-informal} via a reduction from the Subset Product problem \cite{ng2010product}. Interestingly, the fact that the payments are additive across principals while the outcome probabilities are multiplicative means that our reduction has to keep track of both the sum and the product of the numbers involved. To combine both of them into a subset product instance, our reduction works with both the numbers from the subset product instance and the logarithms of those numbers. This results in a more complicated analysis that additionally needs to handle bit-complexity considerations; we defer the details to \cref{sec:computation}.

\paragraph{The gains from cooperation.} Next, we turn to the question of quantifying how much cooperation benefits the principals.\footnote{Note that our model allows each principal to act individually and base her contract only on her own outcome. Thus, cooperation can never hurt their utility.} Namely, what is the largest $\alpha$ for which there are instances where the total utility achieved by $k$ cooperating principals is $\alpha$ times larger than the sum of their utilities when they act individually? At the outset, it is unclear what the bounds on $\alpha$ should look like. In particular, should $\alpha$ depend on the number of principals $k$, the number of actions $n$, the number of outcomes $m$, or perhaps some combination of these?

However, a little thought reveals that $\alpha$ can never be larger than $n$, as $n$ is known to be the largest possible gap between the social welfare and the principal's utility \cite{sim-vs-opt}, and cooperation cannot help the principals surpass the social welfare of the instance. Perhaps surprisingly, we show that this bound is tight, and can be attained with just two principals that are identical to each other.

\begin{theorem}[Informal; formal version in \cref{thm:twoprincipals}]
\label{thm:twoprincipals-informal}
For all $\epsilon \in (0,1)$, there is a contract design instance with two identical principals where cooperation increases the total utility by a factor of at least $\paren*{ 1 - \epsilon } n$.
\end{theorem}

The foregoing discussion implies that any instance proving \cref{thm:twoprincipals-informal} must necessarily exhibit a similar gap between the social welfare and the principal's utility for each individual principal. In our proof, we construct an instance close to the instance in \cite{sim-vs-opt} to show this result. Since the optimal contract in this instance is monotone, we get that \cref{thm:twoprincipals-informal} holds even when restricted to monotone contracts.\footnote{Another well-studied class of contracts is linear contracts. When contracts are restricted to be in this class, the principals get no gains from cooperation, simply because linear contracts can be implemented without cooperation.}

Our next result establishes that these gains from cooperation are not just a worst case phenomenon. We show that, regardless of the instance, having a large enough number of identical principals with that instance gives the principals a total utility arbitrarily close to the social welfare. 

\begin{theorem}[Informal; formal version in \cref{thm:manyprincipals}]
\label{thm:manyprincipals-informal}
For all $\epsilon \in (0,1)$ and all instances, having sufficiently many identical principals with the instance gives them total utility within a factor of $\paren*{ 1 - \epsilon }$ of the social welfare.
\end{theorem}

Roughly speaking, we prove \cref{thm:manyprincipals-informal} by showing that having many identical cooperating principals allows them to sample many outcomes for any action taken by the agent from the individual instance. We use these samples to show that the principals can control the action taken fairly accurately with very little payment, allowing them to force the agent to take the social-welfare maximizing action without paying him more than an $\epsilon$-fraction of the social welfare. Our proof additionally shows that this is possible via payments that have a very simple and succinct description, even though the agent gets non-zero payment on an exponentially large set of outcomes.

Combining \cref{thm:identical-polytime-informal,thm:manyprincipals-informal} shows an interesting ``efficiency-vs-optimality'' result. While \cref{thm:identical-polytime-informal} shows a $k^{ O\paren*{ nm + n^2 } } \cdot \poly(L)$ time bound for computing the optimal contract, the proof of \cref{thm:manyprincipals-informal} shows that a contract that is $\epsilon$-close to the social welfare, and therefore, also $\epsilon$-close to the optimal can be described succinctly in time linear in the size of individual instances, and independent of $k$. We refer the reader to \cref{sec:manyprincipals} for details. 

\paragraph{The power of minimal cooperation.} Our results so far consider cooperating principals that have complete information about each other's outcomes, or principals that do not cooperate at all. Our model of cooperation also allows a more fine-grained perspective, where we can quantitatively control the amount of cooperation between the principals. 

Specifically, in an instance with multiple principals, we say that a contract offered by the principals is separable if the payment by each principal depends only on her own outcomes. Note that separable contracts correspond precisely to the case where there is no cooperation between the principals. We also define two classes of nearly-separable contracts: additively separable and multiplicatively separable. These take a ``cooperation parameter'' $\gamma$ as input and require that any principal's payment can only change by at most $\gamma$ additively (respectively, multiplicatively) based on the outcomes for the other principals. 

We show that these two notions imply very different bounds on the gains possible from cooperation. Although additively separable contracts result in gains bounded by (a function of the) cooperation parameter, multiplicatively-separable contracts can also yield the gains in \cref{thm:twoprincipals-informal} even for small $\gamma$.

\begin{theorem}[Informal; formal version in \cref{thm:additive,thm:multiplicative}]
\label{thm:separable-informal}
\cref{thm:twoprincipals-informal} holds even when the principals are restricted to offer a $\gamma$-multiplicatively separable contract, for any constant $\gamma$. In contrast, when restricted to offer $\gamma$-additively separable contracts, their total utility in any instance is close (parameterized by $\gamma$) to the utility without cooperation.
\end{theorem}

\subsection{Related Work}

\paragraph{Contract design.} The theory of contract design has a rich literature in both economics and computer science. This multifaceted study consists of works on optimizing the principal's utility, approximating the optimal utility via simple contracts, combinatorial contracts, contracts with multiple agents/principals, typed agents, machine learning for contracts, and many more. A non-exhaustive list of works is \cite{GrossmanH83,HM87,Inn90,BabaioffFNW12,EF12,HSV14,robust-linear,sim-vs-opt,roughgarden,ADTC21,GSW21,CMG22,lavi,mul-ag-des,DuttingEFK23,DFPS24,DFGTR24,DEFK25,mul-ag-comb,ACCELTC25}. See \cite{DFT-survey} for an excellent survey.

\paragraph{Common agency.} While we initiate the study of cooperating principals, our work is not the first to study contracts with multiple principals. The classical model for contracts with multiple principals and a single agent is the model of common agency \cite{bernheim-1}. Subsequent work on this model includes \cite{siqueira,prat,lavi,alon}, among others. The main difference between common agency and our model is that common agency restricts the agent to take a single action that applies to all the principals. This naturally sets up competition between the principals, since they may want to incentivize different actions. In contrast, for cooperating principals who want to maximize their total utility, the problem can be reduced to a single principal problem with an aggregated objective. Nevertheless, \cite{siqueira} studies a model with limited cooperation, which is otherwise very different from ours. For instance, in \cite{siqueira}, the cooperating principals can be worse off than when acting independently.

Recent work studies the role of information sharing between principals in a variant of the common agency model~\cite{LimaMR26}, where the agent chooses an action for each principal, and outcomes are independent for each principal. This very closely resembles our own setting, the difference being that Lima et al. study equilibria with two non-cooperative principals, while our work studies cooperating principals. For two principals, actions, and outcomes, their work shows that each principal prefers to share information about their outcome with the other principal. The Pareto-optimal equilibrium with information alliance is exactly the optimal outcome with two cooperating principals in our model (see~\Cref{exm:2}).

\paragraph{Single-agent multitask formulations.} Prior work has also studied contract design settings where the agent can take a subset of actions. These include the multitask analysis of \cite{HM91} and the combinatorial contracts of \cite{DEFK25}. \cite{HM91} model the action of the agent as a continuous vector of efforts while \cite{DEFK25} give the agent a set of discrete actions and allow him to pick any subset of actions. The subset $S$ of actions taken by the agent determines the principal's success probability via a function $f(S)$. Notably, the fact that these works consider only one principal fundamentally distinguishes their setting from ours that is aimed at studying cooperation among multiple principals.

\paragraph{Multi-agent contract design.} The setting of combinatorial agency, where the principal contracts with a set of agents has also been studied \cite{BFN06,BFN10,BabaioffFNW12,EF12,DuttingEFK23,mul-ag-comb}. This setting consists of multiple agents, each of which can either exert effort or shirk. The success probability of the principal is determined by the set of agents who chose to exert effort. More recently, \cite{mul-ag-des,GHC24}, studied a variant of this model where the principal can observe agent-specific outcomes. However, these approaches are, in some sense, ``dual'' to our setting which has many principals interacting with one agent. 

\paragraph{Multi-project contracts.} Very recently, the work of \cite{ACCELTC25} generalized multi-agent contract design to a setting with multiple projects. Here, a principal interacts with a set of agents in order to accomplish multiple projects. These agents can decide whether or not to exert effort and can also decide in which project their effort goes. The utility of the principals is additive across projects and is determined by which agents exert effort towards which project, less any payments made. Although interesting, this work also studies the behavior of multiple interacting agents, instead of studying the cooperation among multiple principals, which is the focus of the current work.

\paragraph{Extensions of our model.} Our model for cooperating principals opens several interesting directions for future research. For example, we allow the agent to take any tuple of actions with one action for each principal. An agent may however only be able to exert limited total effort, and an interesting model would be to consider an ``effort-budget'' that disallows certain tuples that require high effort from the agent. Other possible extensions include considering multiple agents, hierarchical principals, and studying privacy considerations among principals. Our model can also be naturally adapted to study competing principals.

\section{Model and Preliminaries}
\label{sec:prelim}

We start with describing the hidden-action principal-agent model for a single principal. Here, a contract design instance is defined by a tuple $\mathcal{I} = \paren*{ n, m, P, r, c }$ where: \begin{inparaenum}[(1)] \item $n > 0$ is the number of actions available to the agent. We identify these actions with the set $[n]$ and use $a, a'$ {\em etc.} to denote a generic action. \item $m > 0$ is the number of outcomes possible, and we identify these outcomes with the set $[m]$ and use $o, o'$ {\em etc.} for a generic outcome. \item $P \in \mathbb{R}^{ n \times m }$ is the probability matrix with rows corresponding to actions and columns corresponding to outcomes.\footnote{This is also called the \emph{production function}.} Thus, for $a \in [n]$, $p_a$ denotes the $a$th row of $P$ which is a probability distribution over the outcomes $o \in [m]$. Entry $p_a\paren*{ o }$ is the probability of outcome $o$ if the agent chooses action $a$. 
\item $r \in \mathbb{R}^m$ is a vector of non-negative rewards indexed by outcomes $o \in [m]$, with reward $r_o$ accruing to the principal if outcome $o$ occurs. \item $c \in \mathbb{R}^n$ is a vector of non-negative costs indexed by actions $a \in [n]$, with $c_a$ being the disutility the agent incurs to perform action $a$. \end{inparaenum} To make sure the agent is individually rational, we assume in all our instances that there exists an action that has $0$ cost. Finally, for all actions $a \in [n]$, we will use $\E_a\bracket*{ r }$ to denote the expected reward for the principal when the agent performs action $a$. That is, we have $\E_a\bracket*{ r } = \E_{ o \sim p_a }\bracket*{ r_o } = \sum_{ o = 1 }^m p_a\paren*{ o } \cdot r_o$.

\vspace{0.1in}

\noindent \textbf{Contracts.}  To incentivize the agent towards actions with higher expected reward, the principal offers the agent a contract 
$t \in \mathbb{R}^m_+$.\footnote{The nonnegativity constraint is termed \emph{limited liability}, and rules out trivial solutions (e.g.,~\cite{DFT-survey}).} For outcome $o \in [m]$, $t_o$ is the transfer (or the payment) the principal makes to the agent if outcome $o$ occurs. Analogous to our notation for the rewards, we define $\E_a\bracket*{ t } = \sum_{ o = 1 }^m p_a\paren*{ o } \cdot t_o$ as the expected transfer to the agent when he performs action $a$. With this transfer, for all $a \in [n]$, the utilities of the principal and the agent when the agent performs action $a$ are (respectively):
\begin{equation}
\label{eq:util}
\utilpr_a\paren*{ t } = \E_a\bracket*{ r } - \E_a\bracket*{ t } , \hspace{1cm}\text{and}\hspace{1cm} \utilag_a\paren*{ t } = \E_a\bracket*{ t } - c_a .
\end{equation}
Observe that the sum of these utilities is independent of $t$ and is called the social welfare from action $a$. We denote this using $\sw_a = \E_a\bracket*{ r } - c_a$. The maximum social welfare from any action is called the social welfare of the instance,
\begin{equation}
\label{eq:sw}
\sw\paren*{ \mathcal{I} } = \max_{ a \in [n] } \sw_a = \max_{ a \in [n] } \paren*{ \E_a\bracket*{ r } - c_a } .
\end{equation}

\noindent \textbf{Behavior of the parties.} Consider what happens in a contract design instance $\mathcal{I} = \paren*{ n, m, P, r, c }$ when the principal offers a contract $t$. With this contract, the agent will perform an action that maximizes his utility. That is, he performs the action $a^*\paren*{ t } = \argmax_{ a \in [n] } \utilag_a\paren*{ t }$. We say that contract $t$ then incentivizes or implements action $a^*\paren*{ t }$. We assume that ties are broken in favor of the principal's utility. With the agent performing action $a^*\paren*{ t }$, the principal gets a utility of $\utilpr_{ a^*\paren*{ t } }\paren*{ t }$. We denote this simply as $\utilpr\paren*{ t }$. The goal of the principal is to find the optimal contract, {\em i.e.,} the contract that maximizes $\utilpr\paren*{ t }$ over all contracts $t$. We define $\opt\paren*{ \mathcal{I} } = \max_t \utilpr\paren*{ t }$. We say an action is an optimal action if it is implemented by an optimal contract.
Observe that the fact that the agent has a zero cost action implies that he always gets non-negative utility with any contract, which in turn implies that $\opt\paren*{ \mathcal{I} } \leq \sw\paren*{ \mathcal{I} }$ for any instance $\mathcal{I}$.


The following proposition will be useful in our proofs. The proposition is implicit in the proof of Theorem 2 in \cite{sim-vs-opt}, and is also explicitly mentioned in Table 2 of Chapter 4 in \cite{DFT-survey}.

 \begin{restatable}[]{proposition}{lbopt}
 \label{prop:lbopt}
    For any contract design instance $\mathcal{I}$, we have $\sw(\mathcal{I}) \leq n\cdot\opt(\mathcal{I})$, where $n$ is the number of actions.
\end{restatable}

\subsection{Contracts with Multiple Principals}

We now incorporate multiple principals into the hidden-action principal-agent model. Throughout, we use $k$ to denote the number of principals and $\mathcal{I} = \mathcal{I}_1 \otimes \dots \otimes \mathcal{I}_k$ to denote a joint instance, where, for all principals $i \in [k]$, the instance $\mathcal{I}_i = \paren*{ n_i, m_i, P_i, r_i, c_i }$ is a contract design instance as above between the $i$-th principal and the agent. The instances all have the same agent who takes an action $a_i \in \bracket*{ n_i }$ for each instance $i$. This amounts to picking an action tuple $A = \paren*{ a_1, \dots, a_k }$. The outcomes for the principals are drawn independently based on this tuple. We assume the agent costs are additive across principals, so that cost for taking an action $A$ equals $c_A = \sum_{ i = 1 }^k c_{ i, a_i }$.

We also assume that outcomes are independent conditional on the action tuple. Thus given action tuple $A$, the probability of the principals seeing an outcome tuple $O = \paren*{ o_1, \dots, o_k }$, where $o_i \in \bracket*{ m_i }$ is an outcome for principal $i$, is $p_A\paren*{ O } = \prod_{ i = 1 }^k p_{ a_i }\paren*{ o_i }$. When the outcome $O$ occurs, the $i$th-principal gets a reward $r_{ i, o_i }$. Thus, the total reward of the principals is $r_O = \sum_{ i = 1 }^k r_{ i, o_i }$. 

Cooperating principals jointly observe the outcome $O$ and offer a joint contract $t$ that is a function mapping all possible outcome tuples to the transfer for that tuple, which is a non-negative real number. We use $t_O$ to denote the transfer corresponding to the outcome tuple $O$. As before, when presented with a contract $t$ from the principals, the agent picks an action tuple that gives him the highest utility, measured as the difference between the expected transfer for that action and its cost. This action tuple then determines the utility of the principals, which is the expected total reward minus the expected transfer to the agent. With this dynamics, we can define the optimal utility $\opt\paren*{ \mathcal{I} }$ and the social welfare $\sw\paren*{ \mathcal{I} }$ similarly to the above.

\paragraph{Welfare and optimal utility for cooperating principals.} Note that cooperating principals can always choose to ignore each other's outcomes and interact with the agent individually. This means that the optimal utility they get out of the joint instance is never lower than the sum of the utilities of the individual instances. This is captured in the following proposition, which also shows that the social welfare of the joint instance is simply the sum of the social welfare of the individual instances.

\begin{proposition}
\label{prop:jointsw}
Let $k > 0$ and consider $k$ principals with instances $\mathcal{I}_1, \dots \mathcal{I}_k$ respectively. Let $\mathcal{I} = \mathcal{I}_1 \otimes \dots \otimes \mathcal{I}_k$. It holds that:
\begin{enumerate}
\item $\opt\paren*{ \mathcal{I} } \geq \sum_{ i = 1 }^k \opt\paren*{ \mathcal{I}_i }$.
\item $\sw\paren*{ \mathcal{I} } = \sum_{ i = 1 }^k \sw\paren*{ \mathcal{I}_i }$.
\end{enumerate}
\end{proposition}
\begin{proof}
For all $i \in [k]$, let $t_i \in \mathbb{R}^{ m_i }_+$ denote the optimal transfer from the instance $i$. We define a transfer $t$ for the joint instance so that for all outcome tuples $O = \paren*{ o_1, \dots, o_k }$, we have that $t_O = \sum_{ i = 1 }^k t_{ i, o_i }$. We show that, for any action tuple $A = \paren*{ a_1, \dots, a_k }$, the utility of the agent when he takes action $A$ is:
\begin{align*}
\utilag_A\paren*{ t } &= \E_A\bracket*{ t } - c_A \\
&= \sum_{ O = \paren*{ o_1, \dots, o_k } } \prod_{ i = 1 }^k p_{ a_i }\paren*{ o_i } \cdot t_O - c_A \\
&= \sum_{ i = 1 }^k \sum_{ o_i } p_{ a_i }\paren*{ o_i } \cdot t_{ i, o_i } - \sum_{ i = 1 }^k c_{ i, a_i } \tag{Definition of $c_A$ and $t_O$} \\
&= \sum_{ i = 1 }^k \paren*{ \sum_{ o_i } p_{ a_i }\paren*{ o_i } \cdot t_{ i, o_i } - c_{ i, a_i } } \\
&= \sum_{ i = 1 }^k \utilag_{ a_i }\paren*{ t_i } .
\end{align*}
This means the utility of the agent when he takes action $A$ can be decomposed as the sum of his utilities in the individual instances. As all we used in the above derivation was that the transfers and the costs are additive across instances, an analogous result holds for the social welfare and the total utility of the principal. The former proves the second part of the lemma. For the first part, note that our decomposition implies that in each instance, the agent will behave as if the principal for that instance used the optimal transfer for the instance. This, together with the latter, implies the first part of the lemma.
\end{proof}

It is easily seen that the inequality in the first part of \cref{prop:jointsw} can be strict, see the following example.
\begin{example}
\label{exm:2}
Consider an instance $\mathcal{I}$ with two actions and two outcomes with the probabilities, rewards, and costs as depicted in \cref{tab:exm:2}. Observe that the optimal contract for $\mathcal{I}$ is the vector $t = \paren*{ 0, 1/3 }$ that offers a transfer of $0$ on outcome $1$ and a transfer of $1/3$ on outcome $2$. With this contract, the principal's expected utility is $1/3$ implying that $\opt\paren*{ \mathcal{I} } = 1/3$. Now, consider the instance $\mathcal{I} \otimes \mathcal{I}$ and consider the contract $t = \paren*{ 0, 0, 0, 4/5 }$ that only makes a transfer on the outcome $\paren*{ 2, 2 }$. Observe that this contract incentivizes action $\paren*{ 2, 2 }$ and gives the principal a utility of $4/5$. This means that $\opt\paren*{ \mathcal{I} \otimes \mathcal{I} } \geq 4/5 > 2 \cdot \opt\paren*{ \mathcal{I} }$. Thus, the expected utility in the joint instance can exceed the sum of their individual utilities.
\end{example}

\begin{table}
\centering
\begin{tabular}{|c|c|c|c|}
\hline
 & \multicolumn{2}{c|}{Outcomes} & \\
 Actions& {\footnotesize $r_1 = 0$} & {\footnotesize $r_2 = 1$} & Cost \\
 \hline
 $1$ & $2/3$  & $1/3$   & $0$ \\
 \hline
 $2$ & $1/2$ & $1/2$ & $1/18$ \\
 \hline
    \multicolumn{4}{c}{{\footnotesize The instance $\mathcal{I}$.}} \\
\end{tabular}
\hfill
\begin{tabular}{| c|c|c|c|c|c | }
\hline 
 & \multicolumn{4}{c|}{Outcomes} & \\
  Actions& {\footnotesize $r_{(1,1)}=0$} & {\footnotesize $r_{(1,2)}=1$} &{\footnotesize $r_{(2,1)}=1$} & {\footnotesize $r_{(2,2)}=2$} &  Cost\\
 \hline

  $(1,1)$   &   $4/9$    &  $2/9$&   $2/9$&  $1/9$&  $0$\tabularnewline
 \hline
  $(1,2)$&  $1/3$  &  $1/3$   &$1/6$&  $1/6$&  $1/18$\tabularnewline
 \hline
  $(2,1)$ & $1/3$ &  $1/6$&  $1/3$&  $1/6$&  $1/18$\tabularnewline
 \hline
  $(2,2)$ & $1/4$ &  $1/4$&  $1/4$&  $1/4$&  $1/9$\tabularnewline
 \hline
    \multicolumn{6}{c}{{\footnotesize The instance $\mathcal{I} \otimes \mathcal{I}$.}} \\
 
\end{tabular}
\vspace{0.5cm}
\caption{\centering The instances $\mathcal{I}$ and $\mathcal{I} \otimes \mathcal{I}$ of \cref{exm:2}.}
    \label{tab:exm:2}
\end{table}

\subsection{Separable and Nearly Separable Contracts.} 

Note that the proof of \cref{prop:jointsw} uses a transfer that is additive across instances. We call such transfers separable, as they result in the agent treating each instance as separate. Formally, for an instance $\mathcal{I} = \mathcal{I}_1 \otimes \dots \otimes \mathcal{I}_k$ as above, a transfer $t$ for the joint instance is separable if there exist transfers $t_i \in \mathbb{R}^{ m_i }_+$ for the individual instances satisfying for all outcome tuples $O = \paren*{ o_1, \dots, o_k }$ that:
\begin{equation}
\label{eq:separable}
t_O = \sum_{ i = 1 }^k t_{ i, o_i } .
\end{equation}

\paragraph{Nearly separable contracts.} We also study contracts that are close to being separable. For this, we consider a cooperation parameter $\gamma$ and define $\gamma$-additively separable and $\gamma$-multiplicatively separable contracts. Specifically, for $\gamma \geq 0$, we say that a joint contract $t$ is $\gamma$-additively separable if there exist transfers $t_i \in \mathbb{R}^{ m_i }_+$ for the individual instances satisfying for all outcome tuples $O = \paren*{ o_1, \dots, o_k }$ that
\begin{equation}
\label{eq:additively-separable}
\abs*{ t_O - \sum_{ i = 1 }^k t_{ i, o_i } } \leq \gamma .
\end{equation}

Similarly, for $\gamma \geq 1$, we say that a joint contract $t$ is $\gamma$-multiplicatively separable if there exist transfers $t_i \in \mathbb{R}^{ m_i }_+$ for the individual instances satisfying for all outcome tuples $O = \paren*{ o_1, \dots, o_k }$ that
\begin{equation}
\label{eq:multiplicatively-separable}
\frac{ 1 }{ \gamma } \leq \frac{ t_O }{ \sum_{ i = 1 }^k t_{ i, o_i } } \leq \gamma .
\end{equation}
Here, we adopt the convention that $\frac{ 0 }{ 0 } = 1$ and for any positive $x$, we have $\frac{ x }{ 0 } = \infty$.

Similar notions of additive and multiplicative approximations are studied in prior work, e.g., for incentive compatibility and optimal utility in contracts~\cite{roughgarden}, Bayesian equilibria in auctions~\cite{cai2014simultaneous}, and Nash equilibria in bimatrix games~\cite{daskalakis2013complexity}.

\section{The Complexity of Cooperation}
\label{sec:computation}

We first give upper and lower bounds on the complexity of computing an optimal joint contract. Let $\hat{\mathcal{I}} = \mathcal{I}^1 \otimes  \ldots\otimes \mathcal{I}^k$. If for the joint instance $\hat{\mathcal{I}}$ the number of instances $k$ is a constant, then the joint instance has polynomial size; it has $\prod_i n_i$ actions and $\prod_i m_i$ outcomes, and an optimal contract can be computed as for a single instance in polynomial time. If however $k$ is not a constant, then this no longer gives us a polynomial-time algorithm. We show first that if each $\mcI^i$ is identical (i.e., $\hat{\mathcal{I}} = \otimes^k \mathcal{I}$), and has a constant number of actions and outcomes, then the optimal contract can be computed in time polynomial in $k$ and $|\mcI|$. The algorithm extensively uses the symmetrical nature of the joint instance to reduce the search space for the optimal action, and then uses the linear programming approach as for a single instance. 

We next show that for non-identical instances, finding the optimal joint contract is NP-hard, even if the number of actions $n$ and outcomes $m$ in each individual instance is three. Our reduction here is from the \textsc{Subset Product} problem~\cite{ng2010product}. While a similar theme is used in prior work (e.g.,~\cite{roughgarden}), our result requires the use of  both the sum and the product structure from the \textsc{Subset Product} problem. In particular, this requires the use of logarithms which introduces irrational numbers in the reduction, which introduces bit-complexity issues. 
We use techniques to suitably approximate irrational numbers, and show that our reduction holds despite this approximation.

\subsection{Polynomial-Time Algorithm for Identical Instances}
\label{sec:exactalgo}

We use $a$, $a'$ to refer to single actions and $o$, $o'$ for single outcomes in the individual instance $\mcI$. We assume that actions and outcomes are ordered, and will use these as indices as well. We will use capital letters $A$, $A'$, and $B$ for joint actions, and $O$ for joint outcomes. 

Fix a joint action $A = (a_1, \ldots, a_k)$ throughout this section. We will define several terms relative to this fixed action $A$. Let $f \in \mathbb{Z}_+^n$ be the \emph{frequency vector} of individual actions in $A$: $f_a = |\{i: A_i = a\}|$, and $\sum_a f_a = k$. Let $S_k$ be the symmetric group of $k$ elements. Then for a permutation $\sigma \in S_k$, we use $\sigma(O)$ and $\sigma(B)$ to denote the outcome and action obtained by permuting the coordinates of $O$ and $B$ according to $\sigma$ respectively. Further, for a contract $t$, we use $\sigma(t)$ to denote the contract $t'$, where for any joint outcome $O$, $t'(O) = t(\sigma(O))$.

Our first result shows that if action $A$ is optimal, then so is $\sigma(A)$ for any permutation $\sigma$. 

\begin{lemma}
    \label{lem:permuteoptimal} For any permutation $\sigma \in S_k$, outcome $O$, action $B$, and contract $t$, (i) $p_{B}(O) = p_{\sigma(B)}(\sigma(O))$, (ii) $\E_B[\sigma(t)] = \E_{\sigma(B)}[t]$, and (iii) if $t$ implements $B$, then $\sigma^{-1}(t)$ implements $\sigma(B)$.
\end{lemma}

\begin{proof}
    For (i), we note that $p_B(O) = \prod_{i=1}^k p_{b_i}(o_i) = \prod_{i=1}^k p_{b_{\sigma(i)}}(o_{\sigma(i)}) = p_{\sigma(B)}(\sigma(O))$. 
    
    For (ii), 
    \begin{align*}
    \E_{B}[\sigma(t)] & = \sum_O p_{B}(O) t(\sigma(O)) = \sum_O p_{\sigma(B)}(\sigma(O)) t(\sigma(O)) = \E_{\sigma(B)}[t] \, .
\end{align*}

    For (iii), it immediately follows from (ii) that $\E_B[t] = \E_{\sigma(B)}[\sigma^{-1}(t)]$. Further, for any action $B'$, $c_{B'} = c_{\sigma^{-1}(B')}$, and since $t$ implements $B$, $\E_B[t] - c_B \ge \E_{\sigma^{-1}(B')}[t] - c_{\sigma^{-1}(B')}$, and hence $\E_{\sigma(B)}[\sigma^{-1}(t)] - c_B \ge \E_{B'}[\sigma^{-1}(t)] - c_{B'}$. Thus contract $\sigma^{-1}(t)$ implements action $\sigma(B)$. 
\end{proof}

\begin{corollary}
    Let $A$ be an optimal action. Then for any $\sigma \in S_k$, action $\sigma(A)$ is optimal as well. 
    \label{cor:alloptimal}
\end{corollary}

\begin{proof}
    We note first that the expected reward is equal for $A$ and $\sigma(A)$, since it does not matter which principal plays each action. Let $t$ be an optimal contract that implements $A$. By \Cref{lem:permuteoptimal}, $\sigma^{-1}(t)$ implements $\sigma(A)$, and $\E_A[t] = \E_{\sigma(A)}[\sigma^{-1}(t)]$. Then $\utilpr_A\paren*{ t } = \utilpr_{\sigma(A)}\paren*{ \sigma^{-1}(t) }$, and hence $\sigma(A)$ is optimal as well.
\end{proof}

This result thus considerably reduces our search space for an optimal action. Any two actions with the same frequency vector are permutations of each other, and by the corollary, either both are optimal or none are. Thus we can just consider each frequency vector $f$, pick any joint action with this frequency vector, and check for optimality. 

To reduce the search space further, we will need to reduce the set of possible deviating actions, as well as the size of the optimal contract. For deviating actions, we define the set of \emph{deviation-count matrices} 

\[
D^\# = \{D \in \mathbb{Z}_+^{n \times n} : \, D \, \1 = f\} \, .
\]

For $D \in D^\#$, we say a joint action $B$ has deviation count $D$ if $D(a,a') = |\{i \in [k]: \, a_i = a, \, b_i = a'\}|$. Note that $|D^\#| \le (k+1)^{n^2}$. We say that joint actions $B$, $B'$ are equivalent if they have the same deviation count.

To reduce the size of the optimal contract, we next define \emph{outcome-count matrices} 

\[
Q^\# = \{Q \in \mathbb{Z}_+^{n \times m} : \, Q \, \1 = f\} \, .
\]

For $Q \in Q^\#$, we say a joint outcome $O$ has outcome count $Q$ if the entry $Q(a,o) = |\{i \in [k]: \, A_i = a, \, O_i = o\}|$. Note that $|Q^\#| \le (k+1)^{nm}$. We say that joint outcomes $O$, $O'$ are equivalent if they have the same outcome count.

We will use the subset of permutations $\stab(A)$ defined as $\{\sigma \in S_k: \, A = \sigma(A)\}$. It is known that $\stab(A)$ is isomorphic to $\prod_a S_{f_a}$. 

\begin{proposition}
    Actions $B$, $B'$ are equivalent if and only if there exists $\sigma \in \stab(A)$ so that $B' = \sigma(B)$. Similarly, outcomes $O$, $O'$ are equivalent if and only if there exists $\sigma \in \stab(A)$ so that $O' = \sigma(O)$.
    \label{prop:equivalent}
\end{proposition}

\begin{proof}
    We show the proposition for $B$ and $B'$. A similar proof works for $O$ and $O'$ as well. For any individual action $a$, let $S(a) = \{i \in [k]: \, a_i = a\}$.
    
    Since $B$, $B'$ are equivalent, they both have deviation count $D$, for some $D \in D^\#$.  Then for each action $a'$, by definition, both $B$ and $B'$ have the same number of occurrences of $a'$ in $S(a)$ (equal to $D(a,a'))$. For each $a$, $a'$, and $i \le D(a,a')$, let $\sigma$ map the $i$th occurrence of $a'$ in $S(a)$ in $B$, to the $i$th occurrence of $a'$ in $S(a)$ in $B'$. Clearly then $\sigma(B) = B'$. Further, $\sigma$ maps $S(a)$ to $S(a)$ for all $a$, and hence $\sigma \in \stab(A)$.

    For the converse, since $\sigma \in \stab(A)$, it maps $S(a)$ to $S(a)$ for all $a$, and hence preserves the deviation counts. Hence $B$ and $B'$ are equivalent.
\end{proof}

We define a contract $t$ for the joint instance to be succinct if, whenever joint outcomes $O$ and $O'$ are equivalent, then $t_O = t_{O'}$. Note that a succinct contract can be described by ${k+1}^{mn}$ values.\footnote{We emphasize that both equivalence and consequently succinctness are defined with respect to the fixed action $A$.}

\begin{proposition}
    For a succinct contract $t$ and two equivalent joint actions $B$, $B'$, $\E_{B}[t] = \E_{B'}[t]$. 
    \label{prop:succinctequivalent}
\end{proposition}

\begin{proof}
Since $B$ and $B'$ are equivalent, there exists $\sigma \in \stab(A)$ so that $B = \sigma(B')$. From \Cref{lem:permuteoptimal}, we know that $\E_{B}[\sigma^{-1}(t)] = \E_{\sigma^{-1}(B)}[t] = \E_{B'}[t]$. Since $\sigma \in \stab(A)$, by~\Cref{prop:equivalent}, outcomes $O$ and  $\sigma^{-1}(O)$ are equivalent for all outcomes $O$. Since $t$ is succinct, it has the same transfer for all equivalent outcomes, and hence $\sigma^{-1}(t) = t$. Thus again from~\Cref{lem:permuteoptimal}, $\E_{B}[t] = \E_{B'}[t]$.
\end{proof}

\begin{lemma}
    Let $t$ be a contract that implements $A$. Then there exists a succinct contract $\hat{t}$ for which $\E_A[t] = \E_A[\hat{t}]$, and that  implements $A$. 
    \label{lem:succinct}
\end{lemma}

\begin{proof}
    Define $\hat{t} = \frac{1}{|\stab(A)|} \sum_{\sigma \in \stab(A)} \sigma(t)$. To see that $\hat{t}$ is succinct, consider any equivalent outcomes $O$, $O'$. By~\Cref{prop:equivalent}, $O' = \sigma'(O)$ for some $\sigma' \in \stab(A)$. Then
    \[
    \hat{t}(O') = \frac{\sum_{\sigma \in \stab(A)} t(\sigma(\sigma'(O)))} {|\stab(A)|} = \frac{\sum_{\sigma \in \stab(A)} t(\sigma(O))}{|\stab(A)|} = \hat{t}(O) \, .
    \]
    By~\Cref{lem:permuteoptimal}, for any $\sigma \in \stab(A)$, $\E_A[\sigma(t)] = \E_{\sigma(A)}[t] = \E_A[t]$. Then summing over all $\sigma \in \stab(A)$, and by definition of $\hat{t}$, $\E_A[t] = \E_A[\hat{t}]$.
    
    We next show that $\hat{t}$ also implements $A$. Fix an action $B$, and note that for any permutation $\sigma \in \stab(A)$, 
    \[
    \E_A[\sigma(t)] - c_A \ge \E_{\sigma(B)}[t] - c_B = \E_B[\sigma(t)] - c_B \, .
    \]
    Summing both sides over all $\sigma \in \stab(A)$ and dividing by $|\stab(A)|$,
    \[
    \E_A[\hat{t}] - c_A \ge \E_B[\hat{t}] - c_B \, ,
    \]
    Contract $\hat{t}$ thus implements action $A$.
\end{proof}

\begin{restatable}[]{proposition}{permanent}
\label{prop:permanent}
    Given a joint action $B$ and outcome count matrix $Q \in Q^\#$, let $O$ be a joint outcome with outcome count $Q$. Define the $k \times k$ matrix $P$ with entries $P_{ij} = p_{b_i}(o_j)$ if $a_i = a_j$, and $P_{ij} = 0$ otherwise. Then the probability of obtaining an outcome with outcome count $Q$ is exactly
    \[
    \frac{\mathrm{perm}(P)}{\prod_{a,o}Q(a,o)!} \, ,
    \]
    where $\mathrm{perm}(P)$ is the permanent of matrix $P$. The probability can be computed in time $k^{O(n^2)}$.
\end{restatable}

We define $P_B(Q)$ to be the probability that action $B$ obtains an outcome with outcome count $Q$.

\begin{theorem}\label{thm:identical-polytime}
Let $\mathcal{I}$ be an instance with $n$ actions and $m$ outcomes and represented by $L$ bits. A succinct optimal contract for $\otimes^k \mathcal{I}$ can be computed in time $k^{O(nm+n^2)}\poly(L)$.
\end{theorem}

\begin{proof}
    Fix a frequency vector $f$, and an action $A$ with frequency vector $f$. We will write an LP that returns the minimum expected payment to implement $A$. For this, for each outcome count matrix $Q \in Q^\#$, we have a variable $t_Q$, denoting the transfer for each outcome $O$ with outcome count $Q$. Then $t$ describes a succinct contract. Further, for each deviation count matrix $D \in D^\#$, let $A_D$ be a joint action with deviation $D$. Our linear program has a constraint for each such joint action $A_D$, $D \in D^\#$.

\medskip
\noindent
\begin{alignat}{2}
    &\min &&
        \sum_{Q \in Q^\#} P_A(Q) t_Q \notag\\
    &\text{subject to}&&
        \sum_{Q \in Q^\#}\bigl(P_A(Q)-P_{A_D}(Q)\bigr)t_Q
        \ge c_A-c_{A_D}
        \qquad\text{for every }D \in D^\#,
        \notag\\
    &&&t_Q\ge 0
        \qquad\text{for every }Q\in Q^\#.
        \notag
\end{alignat}
 For each frequency vector $f$ we choose an action $A$ with frequency vector $f$, and solve this LP. Assuming feasibility, the LP returns a succinct contract $t$ that implements $A$. We then choose the action and contract that maximizes the principal's utility.

To see this works, for action $A$ let $\{t_Q\}_{Q \in Q^\#}$ be the solution returned by the linear program. Then $t$ is a succinct contract, with $t_Q$ the payment for each outcome $O$ with outcome count $Q$. By definition of $P_B(Q)$, the LP returns a succinct contract that minimizes expected payment, and disincentivizes deviation to any of the chosen actions $\{A_D\}_{D \in D^\#}$. But by~\Cref{prop:succinctequivalent}, all actions equivalent to $A_D$ have the same expected transfer for succinct contract $t$, and the same cost. Since we pick one action for each deviation count matrix $D \in D^\#$, contract $t$ is the minimum payment succinct contract that implements $A$.

Conversely, let $t'$ be a contract that implements $A$. Then by~\Cref{lem:succinct} there is a succinct contract $t$ with the same utility that incentivizes $A$, and clearly $t$ satisfies the constraints in the linear program.

Thus for any action $A$ considered, the linear program (if feasible) returns a succinct contract with minimum expected payment that implements $A$. Now let $A^*$ be an optimal action. Then by~\Cref{cor:alloptimal}, every action with the same frequency vector as $A^*$ is optimal as well. Since we iterate over all possible frequency vectors, picking an action for each, our algorithm will pick an action $A'$ with the same frequency vector as $A^*$, and return an optimal succinct contract. 

For the running time, we note that the algorithm needs to solve at most $(k+1)^n$ linear programs, each of which has size at most $k^{O(nm+n^2)}\poly(L)$, and hence completes in time $k^{O(nm+n^2)}\poly(L)$. \Cref{prop:permanent} shows that the probabilities that are coefficients in the linear program can also be computed in polynomial time.
\end{proof}

\subsection{NP-Hardness for Non-Identical Instances}
\label{sec:nphardness}

We are given $k$ principal-agent instances $\mathcal{I}^1,\ldots,\mathcal{I}^k$ and a threshold $Q$. Let $\hat{\mcI}=\mathcal{I}^1\otimes\cdots\otimes \mathcal{I}^k$. Logarithms in this section have base $e$.

\begin{theorem}
\label{thm:non-identical-np-hard}
Determining if $\opt\paren*{\hat{\mcI}}\ge Q$ is NP-hard, even when each instance $\mathcal{I}^i$ has three actions and three outcomes, and there is an optimal contract supported on a single outcome.
\end{theorem}

We give a polynomial-time reduction from \textsc{Subset Product}~\cite{ng2010product}. Let
$w_1,\ldots,w_k\in\mathbb{Z}_{>0}$ and $V\in\mathbb{Z}_{\ge 0}$ be a \textsc{Subset Product} instance where we need to decide if there exists a subset whose product is $V$. We assume all $w_i \ge 2$, since if $w_i=1$ we can leave it out without affecting the product. Define $w(S):=\prod_{i\in S}w_i$ and $W:=w([k])$. 

The individual contract instances are constructed as follows. For every $i\in[k]$, instance $\mathcal{I}^i$ has actions $\mathtt L,\mathtt R, \mathtt Z$, outcomes $\ell,r,\bot$, and costs $c_i(\mathtt L)=c_i(\mathtt R)=1, c_i(\mathtt Z)=0$. The production function is given by
\[
\begin{array}{c|ccc|c}
    & \ell & r & \bot & \text{cost}\\
    \midrule
    \mathtt L & \tfrac12 & 0 & \tfrac12 & 1\\
    \mathtt R & 0 & \tfrac12 & \tfrac12 & 1\\
    \mathtt Z & \tfrac{\lambda_{i,L}}2
               & \tfrac{\lambda_{i,R}}2
               & 1-\tfrac{\lambda_{i,L}+\lambda_{i,R}}2 & 0
\end{array}
\]
\noindent where $\lambda_{i,L} := \frac{1}{8w_i}$, and $\lambda_{i,R}:= \frac{1}{8}$. The rewards are $r_i(\ell)=2M, r_i(r)=2(M+\gamma \log (w_i))$ and, $ r_i(\bot)=0$
where the parameters $M$ and $\gamma$ will be fixed later. The rewards for the outcome $r$ may be irrational. We will show in Appendix~\ref{sec:appnphardness} that we can replace these logarithmic terms with bounded rational approximations without affecting the proof.

Call a joint action $A=(a_1,\ldots,a_k)$ \emph{good} if
$a_i\in\{\mathtt L,\mathtt R\}$ for every $i$.  Such an action encodes the
subset $S_A:=\{i\in[k]:a_i=\mathtt R\}$. This gives a bijection between good joint actions and subsets of $[k]$.
For a good action $A$, define
\[
    \phi(A):=
    \prod_{i\notin S_A}\lambda_{i,L}
    \prod_{i\in S_A}\lambda_{i,R}.
\]

We will now give a tight bound on the expected payment needed to incentivize any good action.

\begin{restatable}[]{lemma}{minimumpayment}    
\label{lem:minimum-payment}
For every good joint action $A$, the minimum expected payment required to
implement $A$ is exactly $\frac{k}{1-\phi(A)}$.
\end{restatable}

\begin{proof}[Proof sketch.] Fix a good joint action $A$. Let $\mathbf Z=(\mathtt Z,\ldots,\mathtt Z)$ be the all-zeros joint action. To show the lower bound on the expected payment, we compare with $\mathbf Z$. For every single instance outcome $o_i$, $p_{\mathtt Z}(o_i)    \ge \lambda_{i,a_i}p_{a_i}(o_i)$. Therefore, for the joint outcome $O$, $p_{\mathbf Z}(O)\ge \phi(A)p_A(O)$. Therefore, for any contract $t$, $\E_{\mathbf Z}[t]\ge \phi(A)\E_A[t]$. Now if $t$ incentivizes $A$, then since the cost of any good joint action is $k$ while the cost of $\mathbf Z$ is zero, $\E_A[t]-k\ge \E_{\mathbf Z}[t] \ge \phi(A)\E_A[t] $. Hence, $\E_A[t]\ge \frac{k}{1-\phi(A)}$. 

For the upper bound, let $O^A$ and $t^A$ respectively be the outcome and contract given by
\[
    O^A_i:=
    \begin{cases}
        \ell,&a_i=\mathtt L,\\
        r,&a_i=\mathtt R.
    \end{cases}
\]
\[
    t^A(O^A):=2^k\frac{k}{1-\phi(A)},
    \qquad
    t^A(O):=0\quad\text{for }o\ne O^A.
\]
For each $i \in [k]$, $p_{a_i}(O^A_i) = 1/2$, and hence $p_A(O^A)=2^{-k}$, and the expected payment of $t^A$ for action $A$ is exactly $\frac{k}{1-\phi(A)}$. We show in the full proof that the contract $t^A$ implements the action $A$.
\end{proof}

\begin{lemma}
    Any optimal action in the joint instance is a good action.
\end{lemma}
\begin{proof}
    Consider a joint action $A'$ where for some $i \in [k]$, $a'_i = \texttt Z$. Then since $p_{a'_i}(\bot) \ge 3/4$, $A'$ has expected total reward at most $\left(k-\frac34\right)(M+\gamma \log W)$.
On the other hand, the good action $(\mathtt R,\ldots,\mathtt R)$ can be
implemented with principal's utility
\[
    kM + \gamma\log W-\frac{k}{1-8^{-k}}
    >\left(k-\frac34\right)(M+\gamma \log W).
\]
Therefore every optimal induced action is good.
\end{proof}

\noindent \textbf{Encoding \textsc{Subset Product}.} 
Let $\rho:=\frac{8^{-k}}{W}$. We may assume that for the target $V$, $2\le V\le W$. For every good joint action $A$, we have
\begin{align*}
    \phi(A)
    &=
    \prod_{i\notin S_A}\frac{1}{8w_i}
    \prod_{i\in S_A}\frac18 \, = \,     \frac{w(S_A)}{8^k W} \, = \,    \rho w(S_A).
\end{align*}
For a good action $A$, let $t^A$ be the minimum payment contract given by Lemma \ref{lem:minimum-payment} that implements $A$. The maximum principal
utility obtainable while implementing $A$ is
\[
    \utilpr_A(t^A)
    =
    kM+\gamma\log\bigl(w(S_A)\bigr)
    -\frac{k}{1-\rho w(S_A)}.
\]

Define, for $x\in[1,W]$,
\[
    H(x):=\frac{k}{1-\rho x} \quad \text{and} \quad     F(x):=\gamma\log x-H(x).
\]
Note that $\utilpr_A(t^A) = kM + F(w(S_A))$. We now choose $\gamma$ so that $F$ is uniquely maximized at $x=V$. The following lemma is obtained by differentiating $F(x)$, observing that the function is strictly concave, and choosing $\gamma$ so that $F(x)$ is maximized at $x=V$.

\begin{restatable}[]{lemma}{differentiate}
\label{lem:differentiate}
    $F(x)$ is strictly concave in $[1,W]$, and choosing $\gamma = \frac{k \rho V}{(1-\rho V)^2}$ ensures that $F(x)$ is uniquely maximized at $x = V$, and further that if $|x-V| \ge 1$ then $F(V) - F(x) \ge \gamma/2W^2$.
\end{restatable}

\begin{proof}[Proof of~\Cref{thm:non-identical-np-hard}]
Given an instance of \textsc{Subset Product}, we construct the contract instance as shown, and choose threshold $Q:=kM+F(V)$ $=kM+\gamma\log V-\frac{k}{1-\rho V}$.

Suppose first that the \textsc{Subset Product} instance is a
yes-instance. Then there exists $S\subseteq[k]$ such that $w(S)=V$. Let $A$ be the corresponding good action. By Lemma \ref{lem:minimum-payment}, $A$ can be implemented with principal utility
\[
    \utilpr_A(t^A)=kM+F(V)=Q \, ,
\]
where $t^A$ is a minimum payment contract that implements $A$. Therefore, $\opt\paren*{\hat{\mcI}}\ge Q$.

Conversely, suppose that the \textsc{Subset Product} instance is a no-instance. By the preceding argument, every optimal induced joint
action is good. Let $A$ be any good action. Since the instance is a no-instance, $w(S_A)\ne V$. Note that $F$ is uniquely maximized at $V$. Therefore, $F\bigl(w(S_A)\bigr)<F(V)$. Hence,
\[
    \utilpr_A(t^A)
    =
    kM+F\bigl(w(S_A)\bigr)
    <
    kM+F(V)
    =
    Q.
\]
Since every optimal induced action is good, it follows that
\[
    \opt\paren*{\hat{\mcI}}<Q.
\]

In~\Cref{sec:appnphardness}, we show that we can approximate each reward $r_i(r) = 2(M + \gamma \log w_i)$ by a rational approximation which allows us to preserve the hardness.
\end{proof}


\section{The Gains from Cooperation}
\label{sec:gains}

Now, we turn to proving \cref{thm:twoprincipals-informal,thm:manyprincipals-informal} that give tight bounds on the utility gains possible from cooperation. These are proved in \cref{sec:twoprincipals,sec:manyprincipals} respectively. In \cref{sec:twoprincipals}, we show that there exists an instance with two identical principals such that acting individually, they get utility close to the lowest possible for any instance due to a result of \cite{sim-vs-opt}. The tables turn completely when they cooperate, now getting almost the entire social welfare. \cref{sec:manyprincipals} shows that this phenomenon is not worst case. Specifically, with enough identical principals, almost the entire social welfare goes to the principals in any instance.

\subsection{Two Principals Extract Full Welfare in the Worst Case}
\label{sec:twoprincipals}

We next turn to bounds on utility of the principals in joint instances. Our first main result obtains nearly tight bounds on how much larger the total utility for cooperating principals can be, compared to the sum of their individual utilities. We show that, for any $n > 1$, there is a single-principal instance with $n$ actions so that if two principals cooperate, their total utility is almost $n$ times more than their total utility from the two individual instances. We also show that this is the best possible, and in fact, the principals obtain almost the entire social welfare of the joint instance, leaving the agent almost nothing.

\begin{restatable}[]{theorem}{twoprincipals}
\label{thm:twoprincipals}
For all $n > 1$ and all small $\epsilon > 0$, there exists a contract design instance $\mathcal{I} = \paren*{ n, m, P, r, c }$ such that:
\[
\opt\paren*{ \mathcal{I} \otimes \mathcal{I} } \geq 2 \cdot \paren*{ 1 - \epsilon } \cdot \sw\paren*{ \mathcal{I} } \hspace{1cm}\text{and}\hspace{1cm} \sw\paren*{ \mathcal{I} } \geq n \cdot \paren*{ 1 - \epsilon } \cdot \opt\paren*{ \mathcal{I} } > 0 .
\]
\end{restatable}

\begin{proof}[Proof sketch.] We first show that, for the instance $\mathcal{I}$, $\sw\paren*{ \mathcal{I} } = n \cdot \epsilon^{n-1}$ (and is obtained by action $1$), while the optimal utility $\opt\paren*{ \mathcal{I} } \le \epsilon^{n-1}/(1-\epsilon)$. The latter is obtained by observing that any contract $t = (t_1, t_2)$ that incentivizes an action $a$ must give the agent at least as much utility as action $a+1$, and thus the transfer $t_2 \ge 1 - \frac{\epsilon^{n-a}}{1-\epsilon}$. This then allows us to obtain the bound on the principal's utility from any action $a$. This proves the second inequality in the theorem. The contract $t = (0,0)$ incentivizes action $n$, and shows that $\opt(\mathcal{I}) > 0$. 

To show the first inequality in the theorem, for the joint instance $\mathcal{I} \otimes \mathcal{I}$, we show that the contract $t$ that pays the agent $2 - 2n\epsilon^{n-1}\cdot(1-\epsilon)$ on the joint outcome $(2,2)$, and $0$ on other outcomes, incentivizes action $(1,1)$, and obtains utility $2n\epsilon^{n-1}\cdot(1-\epsilon)$ for the principal. This is almost $n$ times larger than twice the optimal utility from the instance $\mathcal{I}$, and is also nearly the entire social welfare of the joint instance (since $\sw\paren*{ \mathcal{I} \otimes \mathcal{I} } = 2 \, \sw\paren*{\mathcal{I}}$). 
\end{proof}

\begin{table}[]
    \centering
   \begin{tabular}{|c|c|c|c|}
   \hline
 & \multicolumn{2}{c|}{Outcomes} & \\
 Actions&  $r_1 = 0$ &  $r_2 = 1$ & Cost \\
 \hline
 $1$ & $0$  & $1$   & $1-n\epsilon^{n-1}$ \\
 \hline
 $\vdots$ & $\vdots$ & $\vdots$ & $\vdots$\\
 \hline
 $a$ & $1-\epsilon^{a-1}$ & $\epsilon^{a-1}$ & $\epsilon^{a-1}-(n-a+1)\cdot\epsilon^{n-1}$ \\
 \hline
 $\vdots$ & $\vdots$ & $\vdots$ & $\vdots$\\
 \hline
 $n$ & $1-\epsilon^{n-1}$  & $\epsilon^{n-1}$   & $0$\\
 \hline
    \multicolumn{4}{c}{{\footnotesize The instance $\mathcal{I}$.}} \vspace{0.5cm} \\
   
\end{tabular}

\begin{tabular}{|c|c|c|c|c|c|}
\hline
 & \multicolumn{4}{c|}{Outcomes} & \\
  Actions&  $r_{(1,1)}=0$ &  $r_{(1,2)}=1$ & $r_{(2,1)}=1$ &  $r_{(2,2)}=2$  &  Cost\\
 \hline

  $(1,1)$   &   $0$    &  $0$&   $0$&  $1$&  $2(1-n\epsilon^{n-1})$\tabularnewline
 \hline
   $\vdots$ & $\vdots$ & $\vdots$ & $\vdots$ & $\vdots$ & $\vdots$\\
 \hline
  $(a,a')$ & $(1-\epsilon^{a-1})(1-\epsilon^{a'-1})$ &  $(1-\epsilon^{a-1})\epsilon^{a'-1}$&  $\epsilon^{a-1}(1-\epsilon^{a'-1})$&  $\epsilon^{(a-1)+(a'-1)}$&  $c_a+c_{a'}$\tabularnewline
 \hline
   $\vdots$ & $\vdots$ & $\vdots$ & $\vdots$ & $\vdots$ & $\vdots$\\
 \hline
 $(n,n)$   &   $(1-\epsilon^{n-1})^2$    &  $\epsilon^{n-1}(1-\epsilon^{n-1})$&   $\epsilon^{n-1}(1-\epsilon^{n-1})$&  $\epsilon^{2(n-1)}$&  $0$\tabularnewline
 \hline
    \multicolumn{6}{c}{{\footnotesize The joint instance $\mathcal{I} \otimes \mathcal{I}$.}} \\
 
\end{tabular}
\vspace{0.5cm}
    \caption{\centering The instances $\mathcal{I}$ and $\mathcal{I}\otimes \mathcal{I}$ of Theorem \ref{thm:twoprincipals}}
    \label{tab:main-1}
\end{table}

It is easy to construct instances where the optimal principal utility is just the social welfare, even if there is only one principal. What \Cref{thm:twoprincipals} actually says is that even if the gap between the social welfare and the optimal principal's revenue is close to a factor of $n$ (which is the worst possible by Proposition \ref{prop:lbopt}) in the single principal case, it is possible that this gap almost vanishes if we have two principals. This also means that \Cref{thm:twoprincipals} is tight up to a factor of $\paren*{ 1-  \epsilon }^2$, i.e., $\opt\paren*{ \mathcal{I} \otimes \mathcal{I} } \le 2n \, \opt\paren*{\mathcal{I}}$. 

\vspace{0.1in}

\noindent \textbf{Remark.} It can be verified that in the single-principal instance $\mathcal{I}$, action $1$ is the optimal action, implemented with the contract $t_1 = 0$, $t_2 = 1- \frac{\epsilon^{n-1}}{1-\epsilon}$. Hence in this case, in the optimal contract, the agent gets utility $n \epsilon^{n-1} - \epsilon^{n-1}/(1-\epsilon)$, or nearly $(n-1)\epsilon^{n-1}$. The social welfare in this instance is $n\epsilon^{n-1}$ (Equation~\ref{eqn:swtwo}), hence almost the entire social welfare goes to the agent. In contrast, in the joint instance, the social welfare is $2n\epsilon^{n-1}$, of which the principal gets $2n(1-\epsilon)\epsilon^{n-1}$ (Equation~\ref{eqn:utilprtwo}), leaving just $2n \epsilon^n$ for the agent. Thus, the agent goes from getting practically all of the social welfare in $\mathcal{I}$, to almost none of it, in the joint instance.

\vspace{0.1in}

\noindent \textbf{Monotone and linear contracts.} Two particular classes of contracts widely studied in contract theory are \emph{monotone} and \emph{linear} contracts. The contract in the proof of~\Cref{thm:twoprincipals} is a monotone contract, since the transfer for an outcome increases with the principal's reward for the outcome. Hence even restricted to monotone contracts, principals retain significant power in joint contracts. A contract $t$ is linear if $t_o = \alpha r_o$ for some constant $\alpha$ and every outcome $o$. A linear contract in a joint instance is separable, since the transfer $t_{(o,o')} = \alpha r_{(o,o')} = \alpha r_o + \alpha r_{o'}$, and hence $t$ is the sum of two linear contracts. Hence, by~\Cref{prop:jointsw}, we get that for linear contracts, cooperation does not benefit the principals.


\subsection{Multiple Cooperating Principals Extract Full Welfare}
\label{sec:manyprincipals}

The upshot of \Cref{sec:twoprincipals} is that there are contract-design instances where two principals achieve the highest possible utility, even if one principal got the lowest possible. Now, we show that, regardless of the instance, a sufficiently large number of principals can obtain almost the entire social welfare, leaving practically nothing for the agent. Formally, letting $\otimes^k \,  \mathcal{I} = \underbrace{ \mathcal{I} \otimes \dots \otimes \mathcal{I} }_{ k \text{~times} }$, we obtain the following result, where the constants in the $\mathcal{O}$ notation depend on the parameters in the instance $\mcI$.

\begin{theorem}
\label{thm:manyprincipals}
For all contract design instance $\mathcal{I} = \paren*{ n, m, P, r, c }$, 
\[
\opt\paren*{ \otimes^k \,  \mathcal{I} } \geq k \cdot \sw\paren*{ \mathcal{I} } \cdot \paren*{ 1 - O\paren*{\sqrt{\frac{\log k}{ k }}}}. 
\]
\end{theorem}
\begin{proof}

For all $a \in [n]$, we will use $a^{(k)}$ in this proof to denote the action $\in [n]^k$ each of whose coordinates equals $a$. Similarly, $O \in [m]^k$ denotes a joint outcome in $\otimes^k \,  \mathcal{I}$, and $o_i$ denotes the $i$-th coordinate of $O$. Fix a contract design instance $\mathcal{I} = \paren*{ n, m, P, r, c }$ and fix an arbitrary action $a \in [n]$ such that $\sw_a > 0$. 

The high-level idea of the proof is to identify a subset of outcomes $\mathcal{O}_a \subseteq [m]^k$ such that the action $a^{(k)}$ is very likely to yield an outcome in $\mathcal{O}_a$ while any action ``far'' away from $a^{(k)}$ is very unlikely to yield an outcome in $\mathcal{O}_a$. This is formalized below. Roughly speaking, the set $\mathcal{O}_a \in [m]^k$ that we will take is the set of all outcomes $O$ such that for all $o \in [m]$, the fraction of coordinates of $O$ that equals to $o$ is close to the probability of sampling outcome $o$ from $p_a$. 
\begin{lemma}
\label{lemma:manyprincipals}
For all actions $a \in [n]$, there exists a subset $\mathcal{O}_a \subseteq [m]^k$ such that:
\begin{enumerate}
\item \label{item:manyprincipals:1} $\displaystyle \sum_{ O \in \mathcal{O}_a } p_{ a^{(k)} }\paren*{ O } \geq 1 - \frac{1}{ 10k } , \text{ and } $
\item \label{item:manyprincipals:2} For all actions $A \in [n]^k$ for which there exists $o \in [m]$ satisfying $\abs*{ \sum_{ i = 1 }^k p_{ a_i }\paren*{ o } - k \cdot p_a\paren*{ o } } \geq 2\Delta_k$, we have $\displaystyle \sum_{ O \in \mathcal{O}_a } p_A\paren*{ O } \leq \frac{1}{ 10k } ,$

\end{enumerate}
where, $\Delta_k = \sqrt{\frac{k}{2}\log(20mk)}$.
\end{lemma}
Now let $a$ be the social welfare-maximizing action in $\mcI$. If $c_a = 0$, then the zero contract trivially implements $a^{(k)}$ in the joint instance, and hence in this case  $\opt\paren*{ \otimes^k \,  \mathcal{I} } = k \cdot \sw\paren*{ \mathcal{I} }$. Thus, assume $c_a > 0$. Let $\mathcal{O}_a$ be as promised by \Cref{lemma:manyprincipals} and consider the contract $t$ for $\otimes^k \,  \mathcal{I}$ such that $t_O = 0$ for all $O \notin \mathcal{O}_a$ and $t_O = \tau$ for all $O \in \mathcal{O}_a$, where $\tau = \paren*{ k + 1 } \cdot c_a$. With this contract, we claim that the utility maximizing action cannot be ``far'' away from $a^{(k)}$. Indeed, for any $A \in [n]^k$ satisfying the condition of \Cref{item:manyprincipals:2} of \Cref{lemma:manyprincipals} above, we have:
\begin{align*}
\utilag_A\paren*{ t } &= \tau \cdot \sum_{ O \in \mathcal{O}_a } p_A\paren*{ O } - \sum_{ i = 1 }^k c_{ a_i } \tag{\Cref{eq:util}} \\
&\leq \frac{ \tau }{ 10k } \tag{\Cref{lemma:manyprincipals}, \Cref{item:manyprincipals:2}} \\
&< \tau \cdot \paren*{ 1 - \frac{1}{ 10k } } - k c_a \tag{as $\tau = (k+1)c_a$} \\
&\leq \tau \cdot \sum_{ O \in \mathcal{O}_a } p_{ a^{(k)} }\paren*{ O } - k c_a \tag{\Cref{lemma:manyprincipals}, \Cref{item:manyprincipals:1}} \\
&\leq \utilag_{ a^{(k)} }\paren*{ t } \tag{\Cref{eq:util}} .
\end{align*}
This means that there exists a utility maximizing action $A^* = \paren*{ a^*_1, \dots, a^*_k }$ for the contract $t$ that does not satisfy the condition of \Cref{item:manyprincipals:2} of \Cref{lemma:manyprincipals}. Now we give a lower bound on the expected reward of action $A^*$.

\begin{align}
    \sum_{i\in[k]}\E_{a^*_i}[r] &= \sum_{o\in[m]}r_o\sum_{i\in[k]}p_{a^*_i}(o)\nonumber\\
    &\ge\sum_{o\in[m]}r_o(kp_a(o)-2\Delta_k)\nonumber\\
    &=k\E_a[r] - 2\Delta_k\sum_{o\in[m]}r_o.\label{eq:reward-lb}
\end{align}
The inequality is true because $A^*$ does not satisfy the condition of \Cref{item:manyprincipals:2} of \Cref{lemma:manyprincipals}. Now we give a lower bound on the principal's utility.

\begin{align*}
\opt\paren*{ \otimes^k \,  \mathcal{I} } &\geq \utilpr_{ A^* }\paren*{ t } \\
&= \sum_{ O \in [m]^k } p_{ A^* }\paren*{ O } \cdot \paren*{ \sum_{ i = 1 }^k r_{ o_i } - t_O } \tag{\Cref{eq:util}} \\
&\geq \sum_{ i = 1 }^k \E_{ a_i^* }\bracket*{ r } - \tau \tag{By linearity of expectation and since $t_O \leq \tau$} \\
&\ge k\E_a[r] - 2\Delta_k\sum_{o\in[m]}r_o-(k+1)c_a\tag{Using \ref{eq:reward-lb}}\\
&=k\cdot\sw_a - 2\Delta_k\sum_{o\in[m]}r_o - c_a. \end{align*}
Since $a$ is the social welfare maximizing action,
\begin{align*}
    \opt\paren*{ \otimes^k \,  \mathcal{I} } &\ge k\cdot\sw(\mathcal{I}) - 2\sum_{o\in[m]}r_o\sqrt{\frac{k}{2}\log(20mk)} - c_a\\
    &=k\cdot\sw(\mathcal{I})\left(1-\frac{2\sum_{o\in[m]}r_o}{k\cdot\sw(\mathcal{I})}\sqrt{\frac{k}{2}\log(20mk)} - \frac{c_a}{k\cdot\sw(\mathcal{I})}\right)\\
    &=k \cdot \sw\paren*{ \mathcal{I} } \cdot \paren*{ 1 - O\paren*{\sqrt{\frac{\log k}{ k }}}}.
\end{align*}\end{proof}

Note that the contract in the proof has a simple and concise description, paying $(k+1)c_a$ for any joint outcome in the set $\mathcal{O}_a$ defined in~\Cref{lemma:manyprincipals}. Further, the result is easily extended when there are multiple instances, rather than a single one. 

\begin{corollary}
    \label{cor:multipleinstance}
    For any contract design instances $\mathcal{I}^1$, $\ldots$, $\mathcal{I}^t$, 
\[
\opt\paren*{ \otimes^k \,  \left(\mathcal{I}^1 \otimes \ldots \otimes \mathcal{I}^t\right) } \geq k \cdot \paren*{ 1 - O\paren*{\sqrt{\frac{\log k}{ k }}} } \cdot\sum_{i=1}^t \sw\paren*{\mathcal{I}^i}. 
\]
\end{corollary}

An instance could appear multiple times in $\mathcal{I}^1$, $\ldots$, $\mathcal{I}^t$, and hence the corollary allows the multiplicity of each instance to increase at different rates. The proof of the corollary is simply by defining $\mathcal{I}$ in~\Cref{thm:manyprincipals} as $\mathcal{I}^1 \otimes \ldots \otimes \mathcal{I}^t$, and observing that the social welfare of $\mathcal{I}$ is then the sum of the individual social welfare in each instance $\mathcal{I}^i$.


\section{The Power of Minimal Cooperation}
\label{sec:minimal}

The previous sections show that if principals cooperate, they can extract nearly the entire welfare in the joint instance, with total utility increasing by a factor of almost $n$. \Cref{prop:jointsw} shows however that if we restrict to \emph{separable} contracts in the joint instance, then the total utility of the principals in the joint instance is at most their total utility in the individual instances. Separable contracts represent an extreme when the contract in the joint instance is simply the sum of contracts in the individual instances. Equivalently, in a separable contract, each principal offers her own individual contract in the joint instance, with the transfer dependent solely on her own outcome.

In this section we take a more fine-grained perspective, going beyond these two extremes. We obtain bounds on the principal's utility if we restrict contracts to be \emph{nearly separable}. Near-separability restricts contracts from being heavily dependent on \emph{joint} outcomes. Do the principals lose their utility advantage with this restriction, or does even a small amount of dependence on joint outcomes give the same power to the principals? 

Our results here show a nuanced picture, that depends on whether the distance of a contract from a separable contract is measured additively or multiplicatively. For contracts that are \emph{additively} nearly-separable, principals cannot do much better than separable contracts. However for \emph{multiplicatively} nearly-separable contracts, principals have nearly the same power as with arbitrary joint contracts, and their utility goes up by nearly $n$ as earlier. 

\subsection{Additively Nearly Separable Contracts}
\label{sec:additive}

Suppose $\hat{t}$ is a $\gamma$-additively separable contract. Therefore, there exists contracts $t_i \in \mathbb{R}^{ m_i }_+$, such that for any joint outcome, $O = (o_1,\ldots,o_k)$, $|\hat{t}_{O} - \sum_{i=1}^kt_{i,o_i}|\le \gamma$. The next theorem shows that if we restrict the principals to offer a $\gamma$-additive nearly separable contract, then they are not going to benefit much by cooperation.
\begin{restatable}{theorem}{additive}\label{thm:additive}
    Suppose $\hat{t}$ is a $\gamma$-additively separable contract that incentivizes $A$. Then,
    \[
    \utilpr_{A}(\hat{t}) \le \left(\sqrt{\sum_{i=1}^k\opt(\mathcal{I}^i)}+\sqrt{2\gamma}\right)^2 + \gamma~\le~ \paren*{ 1 + 2\sqrt{\gamma} } \cdot \sum_{i=1}^k\opt(\mathcal{I}^i) + 4 \sqrt{\gamma} + 3\gamma .
    \]
\end{restatable}
To see why the middle expression implies the last expression, we note the following upper bound on the middle expression:
\[
\utilpr_A(\hat t)\le
\begin{cases}
\sum_{i=1}^k\opt(\mathcal{I}^i)+4 \sqrt{\gamma} + 3\gamma, & \text{if } \sum_{i=1}^k\opt(\mathcal{I}^i) \le 2,\\
\paren*{ 1 + 2\sqrt{\gamma} } \cdot \sum_{i=1}^k\opt(\mathcal{I}^i) + 3\gamma, & \text{if } \sum_{i=1}^k\opt(\mathcal{I}^i) > 2
\end{cases} .
\]
Combining, we get the last expression.

Now, we present a proof sketch of \Cref{thm:additive}. The detailed proof can be found in the appendix.
\begin{proof}[Proof sketch]
     Let $t$ be a joint contract such that, for any joint outcome $O = (o_1,\ldots,o_k)$, $t_O =\sum_{i=1}^k t_{i,o_i}$. Observe that $t$ is a separable contract. Moreover, we can show the following:
     \[
     \utilpr_A(t) \ge \utilpr_A(\hat{t}) - \gamma\tag{*}\label{eq:gamma-gap}
     \]
     
     Now, we consider a contract $t^\alpha$ which a convex combination of the contract $t$ and the reward vector $r$, formally, $t^\alpha:=(1-\alpha)t + \alpha r$, where $\alpha\in(0,1)$. It can be shown that the principal's utility under contract $t^\alpha$ is at least $(1-\alpha)\left(\utilpr_A(t)-(2\gamma/\alpha)\right)$. As $t$ and $r$ are both separable, $t^\alpha$ is also separable. Therefore, the principal's utility under contract $t^\alpha$ is upper bounded by $\sum_{i\in[k]}\opt(\mathcal{I}^i)$.\footnote{The proof is similar to \Cref{prop:jointsw}.} Therefore,
     \begin{align}
    &(1-\alpha)\left(\utilpr_{A}(t) -\frac{2\gamma}{\alpha}\right)\le \sum_{i=1}^k\opt(\mathcal{I}^i)&\nonumber\\
    \implies&\utilpr_{A}(t)\le\frac{\sum_{i=1}^k\opt(\mathcal{I}^i)}{1-\alpha} + \frac{2\gamma}{\alpha}\nonumber
\end{align}
Note that, the last inequality is true for any $\alpha\in(0,1)$. So, we choose the following value which minimizes the right hand side of the last inequality.
\[
\alpha := \frac{\sqrt{2\gamma}}{\sqrt{\sum_{i=1}^k\opt(\mathcal{I}^i)}+\sqrt{2\gamma}}
\]
Using \Cref{eq:gamma-gap} and the above value of $\alpha$, we get the desired bound. 

\end{proof}

\subsection{Multiplicatively Nearly Separable Contracts}
\label{sec:multiplicativelyseparable}

We show now that in contrast to the additive case, if the bounds on distance from separable contracts are multiplicative, then in large enough instances the principals do retain their power and can increase their total utility by nearly a factor of $n$, the number of actions. It is possible even with just two principals.

\begin{restatable}[]{theorem}{multiplicative}
\label{thm:multiplicative}
For all $\gamma > 1$ and $\epsilon_0 \in (0,1/4)$, there exists $n_0$ such that for all $n \ge n_0$, there exists an instance $\mathcal{I}$ with $n$ actions and a $\gamma$-multiplicatively separable contract $\hat{t}$ that incentivizes some action $(a,b)$ for the joint instance $\mathcal{I} \otimes \mathcal{I}$ such that,
\[
\utilpr_{(a,b)}(\hat{t}) \;\ge\; 2n(1-\epsilon_0)^2 \cdot \opt(\mathcal{I}) \, .
\]
\end{restatable}

\noindent \emph{Proof sketch.}
    We consider the instance $\mathcal{I}$ described in Table~\ref{tab:main-1}. We choose $\epsilon = \min \{\epsilon_0, (1- 1/\gamma)^2/9\}$ and $n_0 = 1/\epsilon$. Let $h := 2 - 2n\epsilon^{\,n-1}(1-\epsilon)$, and consider the contract $\hat{t}$ for the joint instance $\mathcal{I}\otimes\mathcal{I}$, defined as follows: $\hat{t}_{(1,1)}= 0$, $\hat{t}_{(2,1)} = \hat{t}_{(1,2)} = h/(2\gamma)$, and $\hat{t}_{(2,2)} = h$. Then $\hat{t}$ is a $\gamma$-multiplicatively separable contract, which can be seen by choosing $t_{1,o_1} = t_{2,o_1} = 0$ and $t_{1,o_2} = t_{2,o_2} = h/2$ and verifying the conditions in~\Cref{eq:multiplicatively-separable}.

The proof of~\Cref{thm:twoprincipals} shows that the contract $t$ that pays $h$ for outcome $(2,2)$ and zero otherwise, incentivizes action $(1,1)$, and obtains utility $2n\epsilon^{n-1} \cdot (1-\epsilon)$ for the principal. Now note that for action $(1,1)$, the outcome $(2,2)$ occurs with probability 1. Hence both the agent's utility and the principal's utility for action $(1,1)$ are the same under contracts $\hat{t}$ and contract $t$ (since the expected transfer remains unchanged). In the full proof, we show that for all $n \ge n_0$, the contract $\hat{t}$ 
incentivizes action $(1,1)$ in the joint instance $\mathcal{I}\otimes\mathcal{I}$.~\Cref{thm:twoprincipals} shows that $\opt\paren*{ \mathcal{I} \otimes \mathcal{I} } \geq 2n \cdot \paren*{ 1 - \epsilon }^2 \cdot \opt\paren*{ \mathcal{I} }$, and since the principal's utility is the same for contracts $t$ and $\hat{t}$, the result follows.


\subsection*{AI disclosure} We used ChatGPT 5.6 Sol to assist with algebraic calculations in two parts of the paper:
\begin{enumerate}
\item In the proof of \Cref{thm:manyprincipals}, it was used to optimize the choice of parameters yielding the asymptotic term $O\left(\sqrt{\frac{\log k}{k}}\right)$.

\item In the proof of \Cref{thm:additive}, it was used to optimize the coefficients of $\sum_{i\in[k]}\opt(\mathcal{I}_i)$ and $\gamma$ in the resulting bound.
\end{enumerate}

Claude was used in catching minor typos and errors in notation. The authors verified the correctness and originality of all content including references.

\bibliographystyle{alpha}

\bibliography{refs}

\appendix
\section*{Appendix}
\section{Proofs from~\cref{sec:exactalgo}}
\label{sec:appexactalgo}

\permanent*

\begin{proof}
    Let $B$, $Q$, $O$, and the matrix $P$ be defined as in the proposition statement. We first note that for any $\sigma \not \in \stab(A)$, the term $\prod_i P_{i,\sigma(i)} = 0$, since for some $i$, $a_i \neq a_{\sigma(i)}$ and $P_{i,\sigma(i)} = 0$. Thus, 
    \[
    \mathrm{perm}(P) = \sum_{\sigma} \prod_i P_{i,\sigma(i)} = \sum_{\sigma \in \stab(A)} \prod_i p_{b_i} (o_{\sigma(i)}) \, .
    \]
    To compute the probability of obtaining an outcome with outcome count $Q$, note that by definition, outcomes with outcome count matrix $Q$ are equivalent, and by~\Cref{prop:equivalent} this set is exactly $\{\sigma(O) : \, \sigma \in \stab(A)\}$ for $O$ as chosen. For each $\sigma \in \stab(A)$, there are exactly $\prod_{a,o} Q(a,o)!$ permutations in $\stab(A)$ that generate the same outcome. Thus, the probability of obtaining an outcome equivalent to $O$ is exactly
    \[
    \frac{1}{\prod_{a,o} Q(a,o)!} \sum_{\sigma \in \stab(A)} \prod_{i=1}^k p_{b_i}(o_{\sigma(i)})
    \]
    which is exactly  $\mathrm{perm}(P) / \prod_{a,o} Q(a,o)!$.
    
    For computation, we first show the following claim.
    
    \begin{claim}
        The matrix $P$ has rank at most $n^2$.
    \end{claim}
    \begin{proof}
        Consider any $i$, $i' \in [k]$ so that $a_i = a_{i'}$ and $b_i = b_{i'}$. We will show that the rows $P_i = P_{i'}$. Let $a = a_i = a_{i'}$ and $b = b_i = b_{i'}$. To see that $P_i = P_{i'}$, for any $j \in [k]$, if $a \neq a_j$ then $P_{ij} = P_{i'j} = 0$. Else, $a = a_j$ and $P_{ij} = P_{i'j} = p_b(o_j)$. Thus the matrix $P$ has at most $n^2$ distinct rows.
    \end{proof}

 Barvinok~\cite{barvinok1996two} gives an algorithm for computing the permanent of a matrix of rank $n^2$ in time $k^{O(n^2)}$, giving us our result. We note that each summation term in the permanent computation has $O(kL + k \log k)$ bits (where $L$ is the bit-complexity of the single instance), so exact rational arithmetic in Barvinok's algorithm can be done in polynomial time. 
\end{proof}

\section{Proofs from~\cref{sec:nphardness}}
\label{sec:appnphardness}

\minimumpayment*

\begin{proof} Fix a good joint action $A$. We first show that any contract that implements $A$ must pay at least $\frac{k}{1-\phi(A)}$ in expectation. As before, for a joint action $B$ and joint outcome $O$, we use $p_B(O)$ to denote the probability of $O$ under $B$. For individual actions, outcomes, and instances, we will use subscripts. Let $\mathbf Z=(\mathtt Z,\ldots,\mathtt Z)$ be the all-zeros joint action. For every single instance outcome $o_i$, $p_{\mathtt Z}(o_i)
    \ge \lambda_{i,a_i}p_{a_i}(o_i)$. Therefore, for the joint outcome $O=(o_1,\ldots,o_k)$, $p_{\mathbf Z}(O)\ge \phi(A)p_A(O)$. Therefore, for any contract $t$, $\E_{\mathbf Z}[t]\ge \phi(A)\E_A[t]$. Now if $t$ incentivizes $A$, then since the cost of any good joint action is $k$ while the cost of $\mathbf Z$ is zero,
\[\E_A[t]-k\ge \E_{\mathbf Z}[t] \ge \phi(A)\E_A[t] \, .\] 
Hence, $\E_A[t]\ge \frac{k}{1-\phi(A)}$. 

Now we show that there exists a contract that incentivizes $A$ with the above payment. Let $o^A$
and $t^A$ respectively be the outcome and contract defined by
\[
    o^A_i:=
    \begin{cases}
        \ell,&a_i=\mathtt L,\\
        r,&a_i=\mathtt R.
    \end{cases}
\]
\[
    t^A(o^A):=2^k\frac{k}{1-\phi(A)},
    \qquad
    t^A(o):=0\quad\text{for }o\ne o_A.
\]
For each $i \in [k]$, $p_{a_i}(o^A_i) = 1/2$, and hence $p_A(o^A)=2^{-k}$, and the expected payment of $t^A$ for action $A$ is exactly $\frac{k}{1-\phi(A)}$. Note that $\frac{k}{1-\phi(A)} - k > 0$, and hence the agent has strictly positive utility for playing action $A$.

Any other joint action $A'$ that switches some coordinate (say $i$) of $A$ from $\mathtt L$ to $\mathtt R$
or vice versa receives zero payment, since $p_{a'_i}(o^A_i) = 0$. Thus any such action is strictly worse than $A$.  It remains to consider the actions that for some subset $T \subseteq [k]$, replace each individual action $a_i$ for $i \in T$ by $\mathtt Z$. Let $A'$ be such an action, thus $a'_i = a_i$ for $i \not \in T$, and $a'_i = \mathtt Z$ for $i \in T$. The expected payment under this action is 

\[
\E_{A'}[t^A] = t^A(o^A) \times \left( \prod_{i \in T} p_{\mathtt Z}(o^A_i) \prod_{i \not \in T} p_{a'_i}(o^A_i)\right)  =  t^A(o^A) \times \left( \prod_{i \in T} \lambda_{i,a_i} \prod_{i \in [k]} p_{a_i}(o^A_i) \right) =  t^A(o^A) \times \frac{\phi_T(A)}{2^k}
\]
\noindent where $\phi_T(A)=\prod_{i \in T}\lambda_{i,a_i}$. Thus $\E_{A'}[t^A] = k \frac{\phi_T(A)}{1-\phi(A)}$. The cost of action $A'$ is $k-|T|$. Thus it is enough to show that
\[
    \frac{k}{1-\phi(A)}\bigl(1-\phi_T(A)\bigr)\ge |T|.
    \tag{IC}
\]
For $T=[k]$, this holds with equality because
$\phi_T(A)=\phi(A)$. We will show later that we can choose $M$ so that the principal prefers $A$ to $\mathbf Z$. If $T$ is a proper nonempty subset and $t=|T|$, then
$t\le k-1$ and $\phi_T(A)\le 4^{-t}$. Moreover, $k\ge t+1$ and $4^t\ge t+1$, hence,
\[
        k(1-\phi_T(A))\ge k(1-4^{-t})
    \ge (t+1)\left(1-\frac1{t+1}\right)=t.
\]
Since $\frac{k}{1-\phi(A)} > k$, condition \emph{(IC)} follows.

 We now justify the tie-breaking statement, i.e., that the principal has greater utility for $A$ than $\mathbf Z$. Choose $M:=2k + \gamma W$. For any good action $A$, its expected reward is
\[
    \E_A[r]=kM+\gamma \log(w(S_A)) \ge kM.
\]
 Note that $\lambda_{i,L}+\lambda_{i,R}<\frac14$. Hence, under joint action $\mathbf Z$,
\[
    \E_{\mathbf Z}[r]\le \frac k4(M+\gamma \log (W)) \le \frac{kM}{2} \,.
\]
thus $\E_A[r]-\E_{\mathbf Z}[r]\ge(kM/2)>k^2 \ge 2k$, by choice of $M$. Now we note that the principal's expected payment for action $A$ is exactly $k/(1-\phi(A))$, and since $\phi(A) \le 1/8$, the expected payment is at most $8k/7 \le 2k$. It follows that the principal strictly prefers action $A$, and the tie-breaking selects $A$, completing the proof.
\end{proof}

\differentiate*

\begin{proof}
We observe that
\[
    F'(x)
    =
    \frac{\gamma}{x}
    -
    \frac{k\rho}{(1-\rho x)^2} \, \text{and}
\]
\[
    F''(x)
    =
    -\frac{\gamma}{x^2}
    -
    \frac{2k\rho^2}{(1-\rho x)^3}
    <0\quad \text{ for $x\in[1,W]$} \, .
\]
Therefore, $F$ is strictly concave on $[1,W]$ which implies $F$ has a unique maximum in $[1,W]$.
Set
\[
    \gamma
    := VH'(V)= \frac{k\rho V}{(1-\rho V)^2}.
\]
Notice that $\gamma$ is rational. With this choice,
\[
    F'(V)
    =
    \frac{\gamma}{V}
    -
    \frac{k\rho}{(1-\rho V)^2}
    =0.
\]
Since $F'(V)=0$, it follows that $F$ is
uniquely maximized at $x=V$.

We next show the explicit gap. Since
$x\le W$,
\[
    F''(x)\le -\frac{\gamma}{W^2}.
\]
Strong concavity and $F'(V)=0$ therefore imply
\[
    F(V)-F(x)
    \ge
    \frac{\gamma}{2W^2}(x-V)^2.
\]
Therefore, if $|x - V| \ge 1$,  $F(V)-F(x)\ge   \Delta$, where $\Delta:=\frac{\gamma}{2W^2}$.
\end{proof}

\paragraph*{Rational approximation of the logarithms.} We will use the following result from \cite{brent1976fast}.
\begin{lemma}[Rational approximation of logarithms \cite{brent1976fast}]\label{lem:rational-log}
Given an integer $q\ge 1$ and a rational number
$\eta\in(0,1)$, one can compute, a rational number $\ell_q$ such that
\[
    0\le \log q-\ell_q\le \eta.
\]

in time polynomial in $bits(q)+\log\frac1\eta$. Hence $\ell_q$ has polynomial bit complexity.
\end{lemma}

Now suppose $\ell_i$ is the rational approximation of $\log w_i$ for all $i\in[k]$. Now set $\eta:=\frac{1}{100kW^2}$ and use Lemma~\ref{lem:rational-log} to compute these $\ell_i$. Let $\ell_V$ be the rational approximation of $\log V$. Thus for any good action $A$, the principal's utility,
\[
\widehat \utilpr_A(t^A)
    =
    kM+\gamma\widehat L(S_A)
    -\frac{k}{1-\rho w(S_A)}.
\]
where $t^A$ is the minimum payment contract that implements $A$, and $\widehat L(S_A) = \sum_{i\in S_A}\ell_i$. Also, for every subset $S\subseteq[k]$,
\[
    \log w(S)-\widehat L(S)
    =\sum_{i\in S}\bigl(\log w_i-\ell_i\bigr)
    \le k\eta.
\]
Therefore, for the good action $A$ corresponding to $S_A$,
\begin{align*}
  \utilpr_A(t^A) - \widehat\utilpr_A(t^A)  = kM+F\bigl(w(S_A)\bigr)-\widehat \utilpr_A(t^A)\le \gamma k\eta
     =\frac{\gamma}{100W^2}
     =\frac{\Delta}{50}.
\end{align*}
Now, we define a new threshold $\widehat Q$ for the proof and will show that the proof still works for this new threshold.
\[
    \widehat Q
    :=kM+\gamma\ell_V-\frac{k}{1-\rho V}-\frac{\Delta}{2}.
\]
Since $\log V-\ell_V\le\eta$, we have,
\[
    Q-\frac{\Delta}{50}-\frac{\Delta}{2}
    \le \widehat Q
    \le Q-\frac{\Delta}{2}.
\]
Suppose first that the \textsc{Subset Product} instance is a
yes instance. Then there exists $S\subseteq[k]$ such that $w(S)=V$. Let $A$ be the corresponding good action. By Lemma \ref{lem:minimum-payment}, $A$ can be implemented with principal's utility
\[
    \widehat \utilpr_A(t^A) = kM + \gamma\widehat L(S_A) - \frac{k}{1-\rho V}
    \ge Q-\frac{\Delta}{50}
    >\widehat Q.
\]
Therefore, $\opt\paren*{\hat{\mcI}}\ge\widehat Q$. Conversely, suppose that the \textsc{Subset Product} instance is a no instance. Let $A$ be any good action. Since the instance is a no instance, $w(S_A)\ne V$. Therefore,
\[
    \widehat \utilpr_A(t^A)
    \le kM+F\bigl(w(S_A)\bigr)
    \le Q-\Delta
    <\widehat Q,
\]
Since every optimal induced action is good, it follows that $\opt\paren*{\hat{\mcI}}<\widehat Q$.

\section{Proofs from~\cref{sec:gains}}

\subsection{Proofs from~\cref{sec:twoprincipals}}
\twoprincipals*

\noindent \emph{Proof of \cref{thm:twoprincipals}.}
Fix $n$ and $\epsilon > 0$ and observe that we can assume that $\epsilon$ is a small enough constant. Consider the instance $\mathcal{I}$ with $n$ actions and $2$ outcomes, outcome $1$ having reward $0$ and outcome $2$ having reward $1$, defined as follows: For all actions $a \in [n]$, the outcome distribution $p_a$ is given by $p_a\paren*{ 1 } = 1 - \epsilon^{ a - 1 }$ and $p_a\paren*{ 2 } = \epsilon^{ a - 1 }$, and the cost is given by:
\begin{equation}
\label{eq:ca:twoprincipals}
c_a = \epsilon^{ a - 1 } - \paren*{ n - a + 1 } \cdot \epsilon^{ n - 1 } .
\end{equation}
The instance $\mathcal{I}$ and the joint instance $\mathcal{I} \otimes \mathcal{I}$ are shown in Table~\ref{tab:main-1}. Observe that the costs $c_a$ are non-negative for a small enough $\epsilon$, and that $c_n = 0$. 

We first compute the social welfare of this instance. From \cref{eq:sw}, this is just the maximum social welfare of any action and we have:
\begin{equation}
\label{eqn:swtwo}
\sw\paren*{ \mathcal{I} } = \max_{ a \in [n] } \paren*{ \E_a\bracket*{ r } - c_a }  = \max_{ a \in [n] } \paren*{ \epsilon^{ a - 1 } - c_a } = \max_{ a \in [n] } \paren*{ n - a + 1 } \cdot \epsilon^{ n - 1 } = n \cdot \epsilon^{ n - 1 } .
\end{equation}
Next, note that $\opt\paren*{ \mathcal{I} } > 0$ as the principals gets positive utility by offering the zero contract, for which the agent chooses action $n$ since $c_n = 0$. Now, we upper bound the principal's optimal utility $\opt\paren*{ \mathcal{I} }$ for this instance. For this, we take an arbitrary action $a \in [n]$ and lower bound the transfer the principal needs to give to the agent to incentivize this action. In turn, this provides an upper bound on the principal's utility, as desired. We first deal with the easy case $a = n$. In this case, we have from \cref{eq:util} that for any transfer $t \in \mathbb{R}^2$, it holds that $\utilpr_n\paren*{ t } \leq \E_n\bracket*{ r } = \epsilon^{ n - 1 }$. For an action $a < n$, note that in order for a transfer $t \in \mathbb{R}^2$ to incentivize action $a$, we must have that action $a$ gives the agent more utility than action $a + 1$. This means that, for all transfers $t \in \mathbb{R}^2$ that incentivize action $a$, we have $\utilag_{ a + 1 }\paren*{ t } \leq \utilag_a\paren*{ t }$. From \cref{eq:util}, this gives $\E_{ a + 1 }\bracket*{ t } - c_{ a + 1 } \leq \E_a\bracket*{ t } - c_a$. Plugging in the values of $p_a$ and \cref{eq:ca:twoprincipals}, we get
\[
\epsilon^{ a - 1 } - \epsilon^a \leq \paren*{ \epsilon^{ a - 1 } - \epsilon^a } \cdot \paren*{ t_2 - t_1 } + \epsilon^{ n - 1 } .
\]
As $t_1 \geq 0$, this means that $t_2 \geq 1 - \frac{ \epsilon^{ n - a } }{ 1 - \epsilon }$. Using this lower bound on the transfer, we can now upper bound the principal's utility. Specifically, for all transfers $t \in \mathbb{R}^2$ that incentivize action $a$, we have from \cref{eq:util} that:
\[
\utilpr_a\paren*{ t } \leq p_a\paren*{ 2 } \cdot \paren*{ 1 - t_2 } \leq \epsilon^{a-1} \cdot \frac{\epsilon^{n-a}}{1-\epsilon} \le \frac{ \epsilon^{ n - 1 } }{ 1 - \epsilon } . 
\]
Combining, we get that $\opt\paren*{ \mathcal{I} } \leq \frac{ \epsilon^{ n - 1 } }{ 1 - \epsilon }$. Since $\sw\paren*{ \mathcal{I} } = n \cdot \epsilon^{ n - 1 }$, we get the second part of the theorem. 

For the first part, we lower bound $\opt\paren*{ \mathcal{I} \otimes \mathcal{I} }$ by giving a contract that gives the principal a high utility. Recall that the outcomes in $\mathcal{I} \otimes \mathcal{I}$ correspond to pairs of outcomes in $\mathcal{I}$ and define the transfer $t$ for $\mathcal{I} \otimes \mathcal{I}$ such that $t_{ \paren*{ 1, 1 } } = t_{ \paren*{ 1, 2 } } = t_{ \paren*{ 2, 1 } } = 0$ and:
\begin{equation}
\label{eq:t22:twoprincipals}
t_{ \paren*{ 2, 2 } } = 2 \cdot \paren*{ 1 - n \cdot \epsilon^{ n - 1 } + n \cdot \epsilon^n } = 2 - 2n\epsilon^{n-1} \cdot \paren*{1-\epsilon}.
\end{equation}
We claim that it suffices to show that the transfer $t$ incentivizes the action $\paren*{ 1, 1 }$ in the instance $\mathcal{I} \otimes \mathcal{I}$. Indeed, assuming this claim is true,  
\begin{equation}
\label{eqn:utilprtwo}
    \opt\paren*{ \mathcal{I} \otimes \mathcal{I} } \geq \utilpr_{ \paren*{ 1, 1 } }\paren*{ t } = 2 - t_{ \paren*{ 2, 2 } } = 2n \cdot \epsilon^{ n - 1 } \cdot \paren*{ 1 - \epsilon }
\end{equation} 
by \cref{eq:util}. Again, since $\sw\paren*{ \mathcal{I} } = n \cdot \epsilon^{ n - 1 }$, we are done. 

To prove the claim, we have to show that for any action pair $\paren*{ a, a' }$, we have $\utilag_{ \paren*{ a, a' } }\paren*{ t } \leq \utilag_{ \paren*{ 1, 1 } }\paren*{ t }$. For $\paren*{ a, a' } = (1,1)$, we obtain 

\begin{equation}
    \label{eqn:utilagtwo}
    \utilag_{ \paren*{1,1} }\paren*{ t } = t_{\paren*{2,2}} - \paren*{2 - 2n\epsilon^{n-1}} = 2n\epsilon^n \, .
\end{equation}

For $\paren*{ a, a' } = (n,n)$, 

\[
\utilag_{ \paren*{n,n} }\paren*{ t } = \epsilon^{2(n-1)} \cdot t_{\paren*{2,2}} = \epsilon^{2(n-1)} \cdot \paren*{2-2n\epsilon^{n-1}(1-\epsilon)} \le 2 \epsilon^{2(n-1)} 
\]
and hence the claimed inequality holds for $\paren*{ a, a' } = (n,n)$. We now consider the case where $\paren*{ a, a' } \not \in \set*{ \paren*{ 1, 1 }, \paren*{ n, n } }$, and in particular, $3 \le a+ a' \le 2n-1$. For this case, we have
\begin{align*}
\utilag_{ \paren*{ a, a' } }\paren*{ t } &= t_{ \paren*{ 2, 2 } } \cdot \epsilon^{ a + a' - 2 } - c_a - c_{ a' } \tag{\cref{eq:util}} \\
&= t_{ \paren*{ 2, 2 } } \cdot \epsilon^{ a + a' - 2 } - \epsilon^{ a - 1 } - \epsilon^{ a' - 1 } + \paren*{ 2n - a - a' + 2 } \cdot \epsilon^{ n - 1 } \tag{\cref{eq:ca:twoprincipals}} \\
&\leq 2 \epsilon^{ a + a' - 2 } - 2 \epsilon^{ \frac{ a + a' - 2}{2} } + \paren*{ 2n - a - a' + 2 } \cdot \epsilon^{ n - 1 } \tag{Convexity of $\epsilon^x$ and $t_{ \paren*{ 2, 2 } } \leq 2$} \\
&\leq - \epsilon^{ \frac{ a + a' - 2}{2} } + \paren*{ 2n - a - a' + 2 } \cdot \epsilon^{ n - 1 } \tag{As $a + a' \geq 3$ and $\epsilon$ is small enough} \\
&\leq \paren*{ 2n - a - a' + 2 - \epsilon^{ \frac{ a + a' - 2n }{2} } } \cdot \epsilon^{ n - 1 } .
\end{align*}
To finish, note that for small enough $\epsilon$, we have $\epsilon^{ -x } \geq 2 + 2x$ for all $x \geq 0.5$. Using this and the fact that $a + a' \leq 2n - 1$, we get $\utilag_{ \paren*{ a, a' } }\paren*{ t } \leq 0$. As the utility maximizing action has non-negative utility, we are done.

\subsection{Proofs from~\cref{sec:manyprincipals}}

We now prove~\cref{lemma:manyprincipals}. The proof relies on the following concentration inequality. 
\begin{lemma}[Hoeffding's inequality]
\label{lemma:hoeffding}
Suppose that $X_1,\ldots,X_n$ are independent random variables taking
values in $[0,1]$. Let $X:=\sum_{i=1}^n X_i$, and $\mu:=\E[X]$. Then, for every $t\geq 0$,
\begin{alignat}{2}
    \Pr\paren*{X-\mu\geq t}
    &\leq
    \exp\paren*{-\frac{2t^2}{n}},
    \label{eq:hoeffding-one-sided}\\
    \Pr\paren*{\abs*{X-\mu}\geq t}
    &\leq
    2\exp\paren*{-\frac{2t^2}{n}}.
    \label{eq:hoeffding-two-sided}
\end{alignat}

\end{lemma}
Specifically, we use \cref{eq:hoeffding-two-sided} to show the following lemma.
\begin{lemma}
\label{lemma:concentration}
Let $m > 0$ and $q_1, \dots, q_k$ be distributions over $[m]$. Define the distribution $Q$ over $[m]^k$ to be $Q = q_1 \times \dots \times q_k$ and the set $\mathcal{O} \subseteq [m]^k$ as:
\[
\mathcal{O} = \set*{ O \in [m]^k \mid \forall o \in [m] : \abs*{ \sum_{ i = 1 }^k \1\paren*{ o_i = o } - \sum_{ i = 1 }^k q_i\paren*{ o } } < \Delta_k } ,
\]
Where, $\Delta_k = \sqrt{\frac{k}{2}\log(20mk)}$. It holds that:
\[
\sum_{ O \in \mathcal{O} } Q\paren*{ O } \geq 1 - \frac{1}{ 10k } .
\]
\end{lemma}
\begin{proof}
Observe that $\sum_{ O \in \mathcal{O} } Q\paren*{ O } = \Pr_{ O \sim Q }\paren*{ O \in \mathcal{O} }$. Given a joint outcome $O=(o_1,\ldots,o_k)\in[m]^k$ and a single outcome $o\in[m]$, define,
\[
X_o(O) := \sum_{i\in[k]}\1\{o_i = o\}.
\]
Therefore, we have $\E_{O\sim Q}[X_o(O)] = \sum_{i\in k}q_i(o)$. Therefore, by Hoeffding's inequality [\cref{eq:hoeffding-two-sided}], we get,
\begin{align*}
    \Pr_{O\sim Q}\left(\left\vert X_o(O) - \E_{O\sim Q}[X_o(O)]\right\vert\ge\Delta_k\right)&=\Pr_{O\sim Q}\left(\abs*{ \sum_{ i = 1 }^k \1\paren*{ o_i = o } - \sum_{ i = 1 }^k q_i\paren*{ o } } \ge \Delta_k \right)\\
    &\le2\exp\left(-\frac{2\Delta^2_k}{k}\right)\\
    &=2\exp\left(-\log(20mk)\right)\\
    &=\frac{1}{10mk}
\end{align*}
Using union-bound over all $o\in[m]$, we get the desired result.
\end{proof}

We are now ready to prove \cref{lemma:manyprincipals}.
\begin{proof}
Fix an arbitrary action $a \in [n]$ and let $\mathcal{O}_a$ be the set from \cref{lemma:concentration} when $q_1 = \dots = q_k = p_a$. \cref{item:manyprincipals:1} follows directly from \cref{lemma:concentration}. For \cref{item:manyprincipals:2}, fix an arbitrary $A \in [n]^k$ and let $o \in [m]$ be as promised by the lemma statement. Observe that it suffices to show that $\mathcal{O}_a$ and $\mathcal{O}_A$ are disjoint where $\mathcal{O}_A$ is the set from \cref{lemma:concentration} when $q_i = p_{ a_i }$ for all $i \in [k]$. For this, suppose for contradiction that $O \in \mathcal{O}_a \cap \mathcal{O}_A$. Using the triangle inequality, this means that:
\[
\abs*{ \sum_{ i = 1 }^k p_{ a_i }\paren*{ o } - k \cdot p_a\paren*{ o } } \leq \abs*{ \sum_{ i = 1 }^k \1\paren*{ o_i = o } - k \cdot p_a\paren*{ o } } + \abs*{ \sum_{ i = 1 }^k \1\paren*{ o_i = o } - \sum_{ i = 1 }^k p_{ a_ i }\paren*{ o } } < 2 \Delta_k, 
\]
a contradiction to the choice of $o$.
\end{proof}

\section{Proofs from~\cref{sec:minimal}}

\subsection{Proofs from~\cref{sec:additive}}
\additive*
\begin{proof}
    Let $t$ be a joint contract such that, for any joint outcome $O = (o_1,\ldots,o_k)$, $t_O =\sum_{i=1}^k t_{i,o_i}$. Suppose $\hat{t} = t + v$. Therefore, for any outcome $O = (o_1,\ldots,o_k)$, $|v_O|\le\gamma$. As $\hat{t}$, incentivizes $A$, then for any joint action $B$,
    \begin{align}
    &\utilag_{A}(\hat{t})\ge\utilag_B(\hat{t})\nonumber\\
    \implies&\utilag_{A}(t) + \E_{A}[v]\ge\utilag_{B}(t) + \E_{B}[v]\nonumber\\
    \implies&\utilag_{A}(t) + 2\gamma\ge \utilag_{B}(t)\label{eq:gamma-bound}
    \end{align}
Let $t^{\alpha} = (1-\alpha)t + \alpha r$, where $r$ is the joint reward vector and $\alpha\in(0,1)$. As $t$ and $r$ both are separable, $t^{\alpha}$ is also separable. Now for any joint action $B$, observe the following:
\begin{align}
    \utilag_B(t^\alpha) &=(1-\alpha)\E_B[t] + \alpha\E_B[r] - c_B\nonumber\\
    &=\utilag_B(t) + \alpha\cdot\utilpr_B(t)\label{eq:alpha-bound}
\end{align}
Now suppose, $t^\alpha$ incentivizes the joint action $B^\alpha$. Therefore,
\begin{align}
    &\utilag_{B^\alpha}(t^\alpha)\ge\utilag_{A}(t^\alpha)&\nonumber\\
    \implies&\utilag_{B^\alpha}(t) + \alpha\cdot\utilpr_{B^\alpha}(t)\ge \utilag_{A}(t) + \alpha\cdot\utilpr_{A}(t)&[\text{using \ref{eq:alpha-bound}}]\nonumber\\
    \implies&\utilag_{A}(t) + 2\gamma + \alpha\cdot\utilpr_{B^\alpha}(t)\ge\utilag_{A}(t) + \alpha\cdot\utilpr_{A}(t)\qquad&[\text{using \ref{eq:gamma-bound}}]\nonumber\\
    \implies&\utilpr_{B^\alpha}(t)\ge \utilpr_{A}(t) - \frac{2\gamma}{\alpha}&\label{eq:gamma-alpha-bound}
\end{align}
Now we will find a lower bound of principal's utility under contract $t^\alpha$.
\begin{align*}
    \utilpr_{B^\alpha}(t^\alpha) &= \E_{B^\alpha}[r - (1-\alpha)t - \alpha\cdot r]&\nonumber\\
    &=(1-\alpha)\E_{B^\alpha}[r-t]&\nonumber\\
    &\ge(1-\alpha)\left(\utilpr_{A}(t) -\frac{2\gamma}{\alpha}\right)&[\text{using \ref{eq:gamma-alpha-bound}}]
\end{align*}
Recall that $t^\alpha$ is a separable contract. Therefore, the principal's utility under $t^\alpha$ can be at most $\sum_{i=1}^k\opt(\mathcal{I}^i)$. Therefore,
\begin{align}
    &(1-\alpha)\left(\utilpr_{A}(t) -\frac{2\gamma}{\alpha}\right)\le \sum_{i=1}^k\opt(\mathcal{I}^i)&\nonumber\\
    \implies&\utilpr_{A}(t)\le\frac{\sum_{i=1}^k\opt(\mathcal{I}^i)}{1-\alpha} + \frac{2\gamma}{\alpha}&\label{eq:all-alpha}
\end{align}
Note that the inequality \ref{eq:all-alpha} is true for any $\alpha\in(0,1)$, so we choose the following value which minimizes the right hand side of the inequality \ref{eq:all-alpha}.
\[
\alpha := \frac{\sqrt{2\gamma}}{\sqrt{\sum_{i=1}^k\opt(\mathcal{I}^i)}+\sqrt{2\gamma}}
\]
With this value of $\alpha$, we conclude the following:
\begin{align}
    &\utilpr_{A}(t)\le\left(\sqrt{\sum_{i=1}^k\opt(\mathcal{I}^i)}+\sqrt{2\gamma}\right)^2&\nonumber\\
    \implies&\utilpr_{A}(\hat{t}) + \E_{A}[v]\le\left(\sqrt{\sum_{i=1}^k\opt(\mathcal{I}^i)}+\sqrt{2\gamma}\right)^2&\nonumber\\
    \implies& \utilpr_{A}(\hat{t}) \le \left(\sqrt{\sum_{i=1}^k\opt(\mathcal{I}^i)}+\sqrt{2\gamma}\right)^2 + \gamma\nonumber
\end{align}

\end{proof}
\subsection{Proofs from~\cref{sec:multiplicativelyseparable}}

\multiplicative*

\begin{proof}
    We consider the instance $\mathcal{I}$ described in Table~\ref{tab:main-1}. We choose $\epsilon = \min \left\{\epsilon_0, (1- 1/\gamma)^2/9\right\}$ and $n_0 = 1/\epsilon$. 
    
Let $h := 2 - 2n\epsilon^{\,n-1}(1-\epsilon)$, and consider the contract $\hat{t}$ for the joint instance $\mathcal{I}\otimes\mathcal{I}$, defined as follows: $\hat{t}_{(1,1)}= 0$, $\hat{t}_{(2,1)} = \hat{t}_{(1,2)} = h/(2\gamma)$, and $\hat{t}_{(2,2)} = h$. Then $\hat{t}$ is a $\gamma$-multiplicatively separable contract, which can be seen by choosing $t_{1,o_1} = t_{2,o_1} = 0$ and $t_{1,o_2} = t_{2,o_2} = h/2$ and verifying the conditions in~\cref{eq:multiplicatively-separable}.

The proof of~\cref{thm:twoprincipals} shows that the contract $t$ that pays $h$ for outcome $(2,2)$ and zero otherwise, incentivizes action $(1,1)$, and obtains utility $2n\epsilon^{n-1} \cdot (1-\epsilon)$ for the principal. Now note that for action $(1,1)$, the outcome $(2,2)$ occurs with probability 1. Hence both the agent's utility and the principal's utility for action $(1,1)$ are the same under contracts $\hat{t}$ and contract $t$ (since the expected transfer remains unchanged). We will now show that for all $n \ge n_0$, the contract $\hat{t}$ 
incentivizes action $(1,1)$ in the joint instance $\mathcal{I}\otimes\mathcal{I}$.~\cref{thm:twoprincipals} shows that $\opt\paren*{ \mathcal{I} \otimes \mathcal{I} } \geq 2n \cdot \paren*{ 1 - \epsilon }^2 \cdot \opt\paren*{ \mathcal{I} }$, and since the principal's utility is the same for contracts $t$ and $\hat{t}$, the result follows.

\begin{claim}\label{clm:lb-epsilon}
    $\utilag_{(1,1)}(\hat{t})\ge\utilag_{(n,n)}(\hat{t})$.
\end{claim}
\begin{proof}
    Note that $\utilag_{(1,1)}(\hat{t}) = 2n\epsilon^{n}$. Now,
    \begin{align*}
        \utilag_{(1,1)}(\hat{t}) - \utilag_{(n,n)}(\hat{t}) &= 2n\epsilon^n - \frac{1}{   \gamma}\cdot \hat{t}_{(2,2)}\cdot\epsilon^{n-1}(1-\epsilon^{n-1}) - \hat{t}_{(2,2)}\cdot\epsilon^{2(n-1)}\\
        &\ge 2n\epsilon^{n} - 2\cdot\frac{1}{\gamma}\cdot\epsilon^{n-1}(1-\epsilon^{n-1}) - 2\epsilon^{2(n-1)}\qquad\tag{$\text{$\hat{t}_{(2,2)}\le 2$}$}\\
        &=2\epsilon^{n-1}\left(n\epsilon - \frac{1}{\gamma} + \epsilon^{n-1}\left(\frac{1}{\gamma}-1\right)\right)> 0\quad\qquad\tag{since, $1\ge\epsilon\ge(1/n)$} 
    \end{align*}
\end{proof}

\begin{claim}\label{clm:ub-epsilon}
 For all actions $(a,a')\notin\{(1,1),(n,n)\}$, $\utilag_{(a,a')}(\hat{t})\le0$ .  
\end{claim}
\begin{proof} For all actions $(a,a')\notin\{(1,1),(n,n)\}$,
    \begin{align*}
\utilag_{ \paren*{ a, a' } }\paren*{ \hat{t} } &= \hat{t}_{ \paren*{ 2, 2 } } \cdot \left(\epsilon^{ a + a' - 2 } + \frac{1}{2\gamma}\left(\epsilon^{a-1}(1-\epsilon^{a'-1})+\epsilon^{a'-1}(1-\epsilon^{a-1})\right)\right) - c_a - c_{ a' } \tag{\cref{eq:util}} \\
&= \hat{t}_{ \paren*{ 2, 2 } } \cdot \left(\epsilon^{ a + a' - 2 } + \frac{1}{2\gamma}\left(\epsilon^{a-1}(1-\epsilon^{a'-1})+\epsilon^{a'-1}(1-\epsilon^{a-1})\right)\right)\\
&\qquad\qquad\qquad- \epsilon^{ a - 1 } - \epsilon^{ a' - 1 } + \paren*{ 2n - a - a' + 2 } \cdot \epsilon^{ n - 1 } \tag{\cref{eq:ca:twoprincipals}} \\
&\leq 2\cdot\left(1-\frac{1}{\gamma}\right)\cdot \epsilon^{ a + a' - 2 } -\left(1-\frac{1}{\gamma}\right)\cdot \epsilon^{   a - 1 }-\left(1-\frac{1}{\gamma}\right)\cdot \epsilon^{   a' - 1 } + \paren*{ 2n - a - a' + 2 } \cdot \epsilon^{ n - 1 }\tag{$\hat{t}_{ \paren*{ 2, 2 } } \leq 2$} \\
&\leq 2\cdot\left(1-\frac{1}{\gamma}\right)\cdot \epsilon^{ a + a' - 2 } - 2\cdot\left(1-\frac{1}{\gamma}\right)\cdot \epsilon^{ \frac{ a + a' - 2}{2} } + \paren*{ 2n - a - a' + 2 } \cdot \epsilon^{ n - 1 } \tag{AM-GM } \\
&\leq - \left(1-\frac{1}{\gamma}\right)\epsilon^{ \frac{ a + a' - 2}{2} } + \paren*{ 2n - a - a' + 2 } \cdot \epsilon^{ n - 1 } \tag{As $a + a' \geq 3$ and $\epsilon\le(1/4)$} \\
&\leq \paren*{ 2n - a - a' + 2 -\left(1-\frac{1}{\gamma}\right) \epsilon^{ \frac{ a + a' - 2n }{2} } } \cdot \epsilon^{ n - 1 } .
\end{align*}

To finish, note that since, $\epsilon\le(1/9)\cdot(1-(1/\gamma))^2$ and $x\ge0.5$, we have $\epsilon^{ -x } \geq (2 + 2x)(1-1/\gamma)^{-1}$. Using this and the fact that $a + a' \leq 2n - 1$, we get $\utilag_{ \paren*{ a, a' } }\paren*{ \hat{t} } \leq 0$. As the utility maximizing action has non-negative utility, we are done.
\end{proof}

From \cref{clm:lb-epsilon,clm:ub-epsilon}, we conclude that if $n\ge n_0 = \max\left\{1/\epsilon_0,9\gamma^2/(\gamma - 1)^2\right\}$, then $\hat{t}$ incentivizes $(1,1)$.
\end{proof}

\end{document}